\documentclass[pre,showpacs,twocolumn,amsmath,amssymb,superscriptaddress,floatfix]{revtex4}

\usepackage[dvipdf]{graphicx}

\usepackage{bm}
\usepackage[justification=centerlast]{caption}

\usepackage{subfig}

\usepackage{latexsym}

\begin{document}


\title{Spatial Rock--Paper--Scissors Models with Inhomogeneous Reaction Rates}

\author{Qian He} \email{heq07@vt.edu}
\affiliation{Department of Physics and
     	Center for Stochastic Processes in Science and Engineering,
     	Virginia Polytechnic Institute and State University,
     	Blacksburg, Virginia 24061-0435, U.S.A.}
\author{Mauro Mobilia}\email{M.Mobilia@leeds.ac.uk}
\affiliation{Department of Applied Mathematics, School of Mathematics,
        University of Leeds, Leeds LS2 9JT, U.K.}
\author{Uwe C. T\"auber}\email{tauber@vt.edu}
\affiliation{Department of Physics and
        Center for Stochastic Processes in Science and Engineering,
        Virginia Polytechnic Institute and State University,
        Blacksburg, Virginia 24061-0435, U.S.A.}

\begin{abstract}
We study several variants of the stochastic four-state rock--paper--scissors 
game or, equivalently, cyclic three-species predator--prey models with 
conserved total particle density, by means of Monte Carlo simulations on one- 
and two-dimensional lattices.
Specifically, we investigate the influence of spatial variability of the 
reaction rates and site occupancy restrictions on the transient oscillations 
of the species densities and on spatial correlation functions in the 
quasi-stationary coexistence state.
For small systems, we also numerically determine the dependence of typical
extinction times on the number of lattice sites.
In stark contrast with two-species stochastic Lotka--Volterra systems, we find
that for our three-species models with cyclic competition quenched disorder in
the reaction rates has very little effect on the dynamics and the long-time
properties of the coexistence state. 
Similarly, we observe that site restriction only has a minor influence on the 
system's dynamical properties.
Our results therefore demonstrate that the features of the spatial
rock--paper--scissors system are remarkably robust with respect to model
variations, and stochastic fluctuations as well as spatial correlations play a
comparatively minor role.
\end{abstract}


\date{\today}

\maketitle

\section{Introduction}
\label{introd}

Understanding the origin of and maintaining biodiversity is of obvious
paramount importance in ecology and biology
\cite{May,Maynard,Michod,Sole,Neal}.
In this context, paradigmatic schematic models of predator--prey interaction
that build on the classic Lotka--Volterra system \cite{Lotka,Volterra} have
been widely studied.
Specifically, systems with cyclic dominance of competing populations have been
suggested to provide a mechanism to promote species diversity; there are also
natural connections to evolutionary game theory
\cite{MaynardSmith,Hofbauer,Nowak,Szabo,Redner,Mobilia}.
A minimal yet non-trivial model for cyclic competition is the three-species
cyclic predator--prey system with standard Lotka--Volterra predation
interactions, essentially equivalent to the familiar rock--paper--scissors
(RPS) game \cite{MaynardSmith,Hofbauer,Nowak,Szabo}.
This RPS system has, for example, been used to model the cyclic competitions
between three subspecies of certain Californian lizards \cite{Sinervo,Zamudio},
and the coevolution of three strains of {\em E. coli} bacteria in microbial
experiments \cite{Kerr}.
Other examples include coral reef invertebrates \cite{Jackson} and overgrowths
by marine sessile organisms \cite{Buss,Burrows}.
In this simple RPS model, one lets `rock' (species $A$) smash `scissors'
(species $B$), `scissors' cut `paper' (species $C$), and `paper' wrap `rock'.
Already for a non-spatial RPS system, the presence of intrinsic stochastic
fluctuations (reaction noise) makes the system eventually evolve to one of the
three extinction states where only one species survives
\cite{Ifti,Reichen,Frean,Berr}.
For example, if the reaction rates in the system are not equal, one
intriguingly observes the `weakest' species, with the smallest predation rate,
to survive, whereas the other two species always die out \cite{Berr,Frean}.
When the model is extended to include spatial degrees of freedom, say by
allowing particles to hop to nearest-neighbor sites on a lattice and interact
upon encounter, spatial fluctuations and correlations further complicate the
picture.
For instance, species extinction still prevails in one-dimensional RPS models
\cite{Tainaka1,Frache1,Frache2,Provata}, but the system settles in a 
coexistence state when the species are efficiently mixed through particle 
exchange (but see also Ref.~\cite{Pleimling}).
In contrast, two-dimensional RPS systems are characterized by coexistence of
the competing species, and the emergence of complex spatio-temporal structures
such as spiral patterns \cite{Frean,Tainaka1,Provata,Tainaka2,Tainaka3,Szabo2,
Perc,Reichenbach1,Reichenbach2,Reichenbach3,Reichenbach4,Tsekouras,Matti}.
Recently, Reichenbach {\em et al.} extensively studied the four-state RPS model
without conservation law
\cite{Reichenbach1,Reichenbach2,Reichenbach3,Reichenbach4}, and it is now
well-established that cyclic reactions in conjunction with diffusive spreading
generate spiral patterns (when the system is sufficiently mixed).
In model variants that incorporate conservation of the total population
density, on the other hand, spiral patterns do not occur
\cite{Tainaka1,Frean,Matti}; also, when the species mobility is drastically
enhanced through fast particle exchange processes, the spiral patterns are
destroyed as well, and the system eventually reaches an extinction state
\cite{Reichenbach1,Matti}.

This work is motivated by the following question: 
Which are the crucial model ingredients to be included in order to attain a 
further degree of realism?
To this end, we carried out Monte Carlo simulation studies on the influence of 
the carrying capacity and environmental inhomogeneity on the properties of a 
class of spatial RPS models where the total population size is conserved 
(zero-sum games) \cite{Hofbauer,Szabo,Ifti,Reichenbach1,Frean,Tainaka1,
Frache1,Frache2,Provata,Tainaka2,Tainaka3,Tsekouras,Matti,Dobrienski}.
In our stochastic lattice models, the carrying capacity (i.e., the maximum 
population size that can be sustained by the environment) is implemented 
through site occupation number restrictions.
Environmental variability is modeled through assigning local reaction rates 
that are treated as quenched random variables drawn from a uniform 
distribution.
Our extensive numerical study shows that carrying capacity and quenched
disorder have little influence on the oscillatory dynamics, spatial correlation
functions, and extinction times in the RPS model system.
This demonstrates a quite remarkable robustness of this class of models. 
From a modeling perspective, this establishes the essential equivalence of 
rather distinct model variants. 
We emphasize that this outcome is nontrivial, as is, for example, revealed by a 
comparison with the two-species Lotka-Volterra system \cite{Ivan,Mark}, where
spatially varying reaction rates may cause more localized clusters of activity
and thereby enhance the fitness of both predator and prey species 
\cite{Ulrich}.

Naturally, lattice models should be viewed as coarse-grained representations of
a metapopulation system where each lattice site or cell can be interpreted as a
`patch' (or `island') populated by a `deme' (or `local community') 
\cite{Hubbell,Hanski}.
For the sake of simplicity (i.e. to try to understand the simplest possible 
systems before venturing further), we here restrict our presentation to RPS 
model variants that obey a conservation law (even though that has no particular
ecological motivation).
As empty lattice sites are allowed, we shall refer to our model system as a 
class of four-state RPS models with conservation law (if all sites are at most 
occupied by a single individual, each of them can be in one of four states).
While the presence or absence of the conservation of the 
total number of particles is crucial for the emergence of spiral waves in RPS 
systems~\cite{Matti}, this is not the case for the properties studied here. 
In fact, it turns out that our conclusions on the effects (or lack thereof) of 
limited carrying capacity and random environmental influences on the transient
population oscillations, spatial correlation functions, and species extinction 
times are common to models both with and without conservation laws \cite{HMT}.

Our paper is structured as follows: In Sec.~II, we define our model, the
stochastic four-state spatial rock--paper--scissors (RPS) game or cyclic
three-species predator--prey system with conservation of total population
density, and briefly review the results obtained from the mean-field rate
equation approximation.
In Sec.~III, we introduce our Monte Carlo simulation algorithm and discuss the
detailed model variants we have explored.
We then present results for the species' time-dependent densities, associated
frequency power spectra, and spatial correlation functions to analyze the
influence of quenched spatial disorder in the reaction rates and site
occupation restriction on the temporal evolution and quasi-stationary states of
this system, both in two dimensions and for a one-dimensional lattice.
We also compare our numerical findings with the mean-field predictions, and
obtain the mean extinction time (for the first species to die out) as function
of system size.
Finally, we provide a summary of our results and concluding remarks.

\section{Model and rate equations}
\label{meanft}

The rock--paper--scissors (RPS) model describes the cyclic competition of three
interacting species that we label $A$, $B$, and $C$.
We consider the following (zero-sum \cite{Hofbauer}) predator--prey type
interactions:
\begin{eqnarray}
  A + B &\to& A + A  \quad {\rm with \ rate} \ k_a \ , \nonumber \\
  B + C &\to& B + B  \quad {\rm with \ rate} \ k_b \ , \label{react} \\
  C + A &\to& C + C  \quad {\rm with \ rate} \ k_c \ . \nonumber
\end{eqnarray}
Note that these irreversible reactions strictly conserve the total number of
particles.
We remark that naturally other variants of the RPS dynamics could also be
considered; notably the four-state May--Leonard model which does not conserve 
the total particle density \cite{MayLeonard} has attracted considerable 
attention, see e.g. Refs.~\cite{Hofbauer,DurrettLevin,Reichenbach2}.
As will be demonstrated elsewhere, the conclusions presented here on the 
effects of carrying capacity and spatial reaction rate variability remain 
essentially unchanged for this system \cite{HMT}.
To generalize the above reaction model to a spatially extended lattice version,
we allow empty sites (as a fourth possible state) and let the reactions happen
only between nearest neighbors.
In addition, we introduce nearest-neighbor particle hopping with rate $D$ (if
at most one particle is allowed per lattice site, this process takes place only
if an adjacent empty site becomes selected at each time step).

Within the mean-field approximation, wherein any correlations and spatial
variations are neglected, the following set of three coupled rate equations for
homogeneous population densities $a(t)$, $b(t)$, and $c(t)$, with fixed total
population density $a(t) + b(t) + c(t) = \rho = {\rm const}$ describes the
system's temporal evolution,
\begin{eqnarray}
  \partial_t \, a(t) &=& a(t) \left[ k_a \, b(t) - k_c \, c(t) \right] \, ,
  \nonumber \\
  \partial_t \, b(t) &=& b(t) \left[ k_b \, c(t) - k_a \, a(t) \right] \, , \\
  \partial_t \, c(t) &=& c(t) \left[ k_c \, a(t) - k_b \, b(t) \right] \, .
  \nonumber
\label{rateq}
\end{eqnarray}
These coupled rate equations possess a reactive fixed point, where all three
species coexist, $(a^*,b^*,c^*) = (k_b,k_c,k_a) \rho/(k_a+k_b+k_c)$, which is
marginally stable (see also Ref.~\cite{Frean}).
Indeed, introducing new variables $\delta a(t) = a(t)-a^*$,
$\delta b(t) = b(t)-b^*,$ $\delta c(t) = c(t)-c^*$, and utilizing the
conservation law $\delta a + \delta b + \delta c = 0$, we may express the first
two rate equations in terms of $\delta a$ and $\delta b$.
Linearizing about the reactive fixed point then gives
\begin{equation}
  \left( \begin{array}{c} \partial_t \, {\delta a} \\ \partial_t \, {\delta b}
  \\ \end{array} \right) = L \left( \begin{array}{c} \delta a \\ \delta b \\
  \end{array} \right) \ ,
\end{equation}
with the linear stability matrix
\begin{equation}
  L = \frac{\rho}{k_a+k_b+k_c} \left( \begin{array}{cc} k_b \, k_c &
  k_b \, (k_a+k_c) \\ -k_c \, (k_a+k_b) & -k_b \, k_c \\ \end{array} \right)\ ,
\end{equation}
with eigenvalues $\lambda = \pm i \, \rho \, \sqrt{k_a \, k_b \, k_c  /
(k_a+k_b+k_c)} = \pm i \omega$, where $f = \omega / 2 \pi$ represents a
characteristic oscillation frequency, e.g., for total density $\rho = 1$ and
$k_a = 0.2$, $k_b = 0.5$, $k_c = 0.8$, $\lambda = \pm i \, 2 \sqrt{3} / 15$,
and the typical frequency is $f \approx 0.037$.
We will use these mean-field values later to compare with the simulation
results.
In the special case of symmetric reaction rates where $k_a = k_b = k_c = k$, we
get $\lambda= \pm i \, k / \sqrt{3}$; for example, if $\rho = 1$ and $k=0.5$,
then $\lambda= \pm i \, \sqrt{3} / 6$ and $f \approx 0.046$.
In addition, the system also has three absorbing states, with only a single
species surviving ultimately: $(\rho,0,0)$, $(0,\rho,0)$, and $(0,0,\rho)$.
Within the mean-field approximation, these fixed points are all linearly
unstable.
However, in any stochastic model realization on a finite lattice, temporal
evolution would ultimately terminate in one of these absorbing states, as we
shall explore for small systems below.

\section{Monte Carlo simulation results}

\subsection{Model variants and quantities of interest}
\label{modint}

We investigate stochastic RPS systems on one- and two-dimensional lattices with
periodic boundary conditions.
At each time step, one individual of any species is selected at random, then
hops to a nearest-neighbor site, if the number of particles on the chosen
target site is empty.
Otherwise, one of the particles on the chosen neighboring site is selected
randomly and undergoes a reaction with the center particle according to the
scheme and rates specified by (\ref{react}) if both particles are different.
The outcome of the reaction then replaces the eliminated particle.
Note that predation reactions always involve neighboring particles; on-site
reactions do not occur.
This has the advantage of permitting us to treat the model variants with and
without site occupation number restriction within the same setup, allowing for
direct comparison.
A similar approach was already adopted for the two-species lattice
Lotka--Volterra model, where we confirmed earlier that nearest-neighbor
predation interactions and strictly on-site reactions lead (without loss of
generality) to essentially identical macroscopic features \cite{Ivan,Mark}.

If the selected and focal particle are of the same species, the center particle
just hops to its chosen neighboring site.
For our model variants with site occupancy restriction, the hopping process
only takes place if the total number of particles on the target site is less
than the maximum occupancy number (local carrying capacity) $n_m$.
In this work, we set $n_m = 1$; i.e., each lattice site can either be empty or
occupied by a single particle of either species $A$, $B$, or $C$ (which gives
four possible states for each site).
Once on average each individual particle in the lattice has had the chance to
react or move, one Monte Carlo step (MCS) is completed; thus the corresponding
simulation time is increased by $\delta t \sim N^{-1}$.
Also note that the hopping processes set the fundamental time scale; basically
the reaction rates are measured in units of the diffusivity $D$ (unless
$D = 0$).

\begin{table}[!t]
\begin{tabular}{|c|c|c|} \hline
  \textbf{Model} & \textbf{Reaction rates} & \textbf{Site restriction} \\
  \hline
  \textbf{1} & homogeneous rate: $k=0.5$ & no restriction \\ \hline
  \textbf{2} & homogeneous rate: $k=0.5$ & at most one particle \\ \hline
  \textbf{3} & uniform rate distribution & no restriction \\ \hline
  \textbf{4} & uniform rate distribution & at most one particle \\ \hline
\end{tabular}
\caption{\label{models} List of stochastic lattice RPS model variants.}
\end{table}
First, we shall study models with uniform symmetric reaction rates
($k_a = k_b = k_c = k = 0.5$); next we simulate systems with quenched spatial
disorder by drawing the reaction probabilities $k$ at each lattice site from a
uniform distribution on the interval $[0,1]$.
Therefore, this distribution has the same mean reaction rate $1/2$ as the
homogeneous rate in the model with fixed reaction rates, allowing for direct
comparison of the relevant numerical quantities.
The four basic different model variants we have investigated are summarized in
Table~\ref{models}.
In addition, we have studied systems with asymmetric reaction rates, both
uniform and subject to quenched randomness with flat distribution.
Besides the time-dependent population densities $a(t)$, $b(t)$, and $c(t)$,
averaged over typically 50 individual simulation runs, we also investigate
their corresponding temporal Fourier transforms
$a(f) = \int a(t) \, e^{2 \pi i f t} \, dt$,
and the equal-time two-point occupation number correlation functions
(cumulants)
$C_{AB}(x,t) = \langle n_A(i+x,t) \, n_B(i,t) \rangle - a(t) \, b(t)$, where
$i$ denotes the site index, and similarly for the other species, as well as
$C_{AA}(x,t)$, etc.
In addition, for small systems with $N$ lattice sites we have numerically
computed the mean extinction time $T_{\rm ex}(N)$ defined as the average time
for the first of the three species to die out~\cite{Schutt}.
For the one-dimensional four-state RPS model, we have also determined the time
evolution of the typical single-species domain size
$\langle \lambda(t) \rangle$, see Sec.~\ref{onedim}.

\subsection{Two-dimensional stochastic RPS lattice models: symmetric rates}
\label{twodim}

\begin{figure}[!t]
\begin{center}
\subfloat[]{\label{fig1:twodim}
\includegraphics[width=0.48\textwidth]{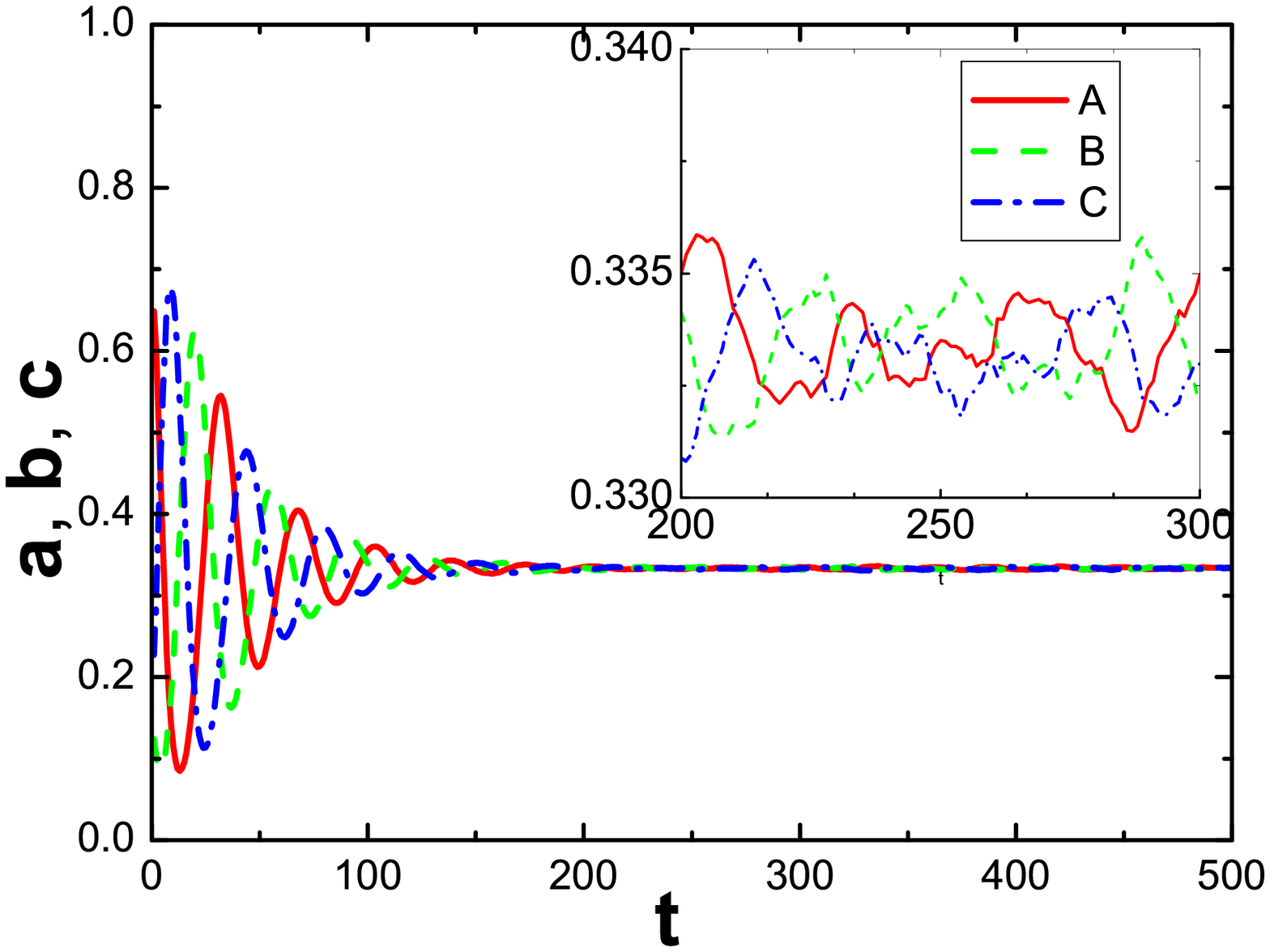}} \smallskip \\
\subfloat[]{\label{fig1:snpsh1}
\includegraphics[width=0.145\textwidth]{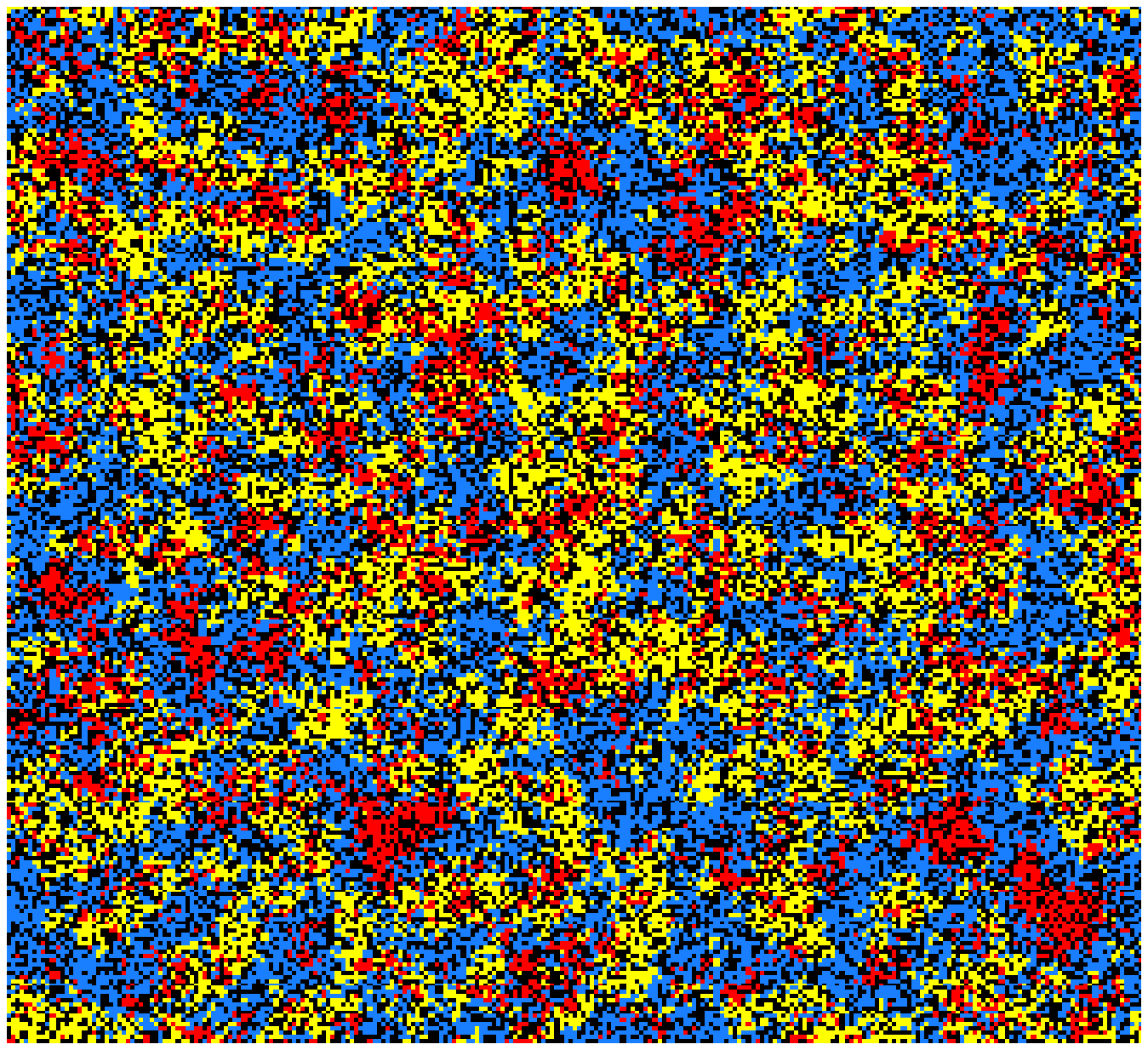}} \
\subfloat[]{\label{fig1:snpsh2}
\includegraphics[width=0.145\textwidth]{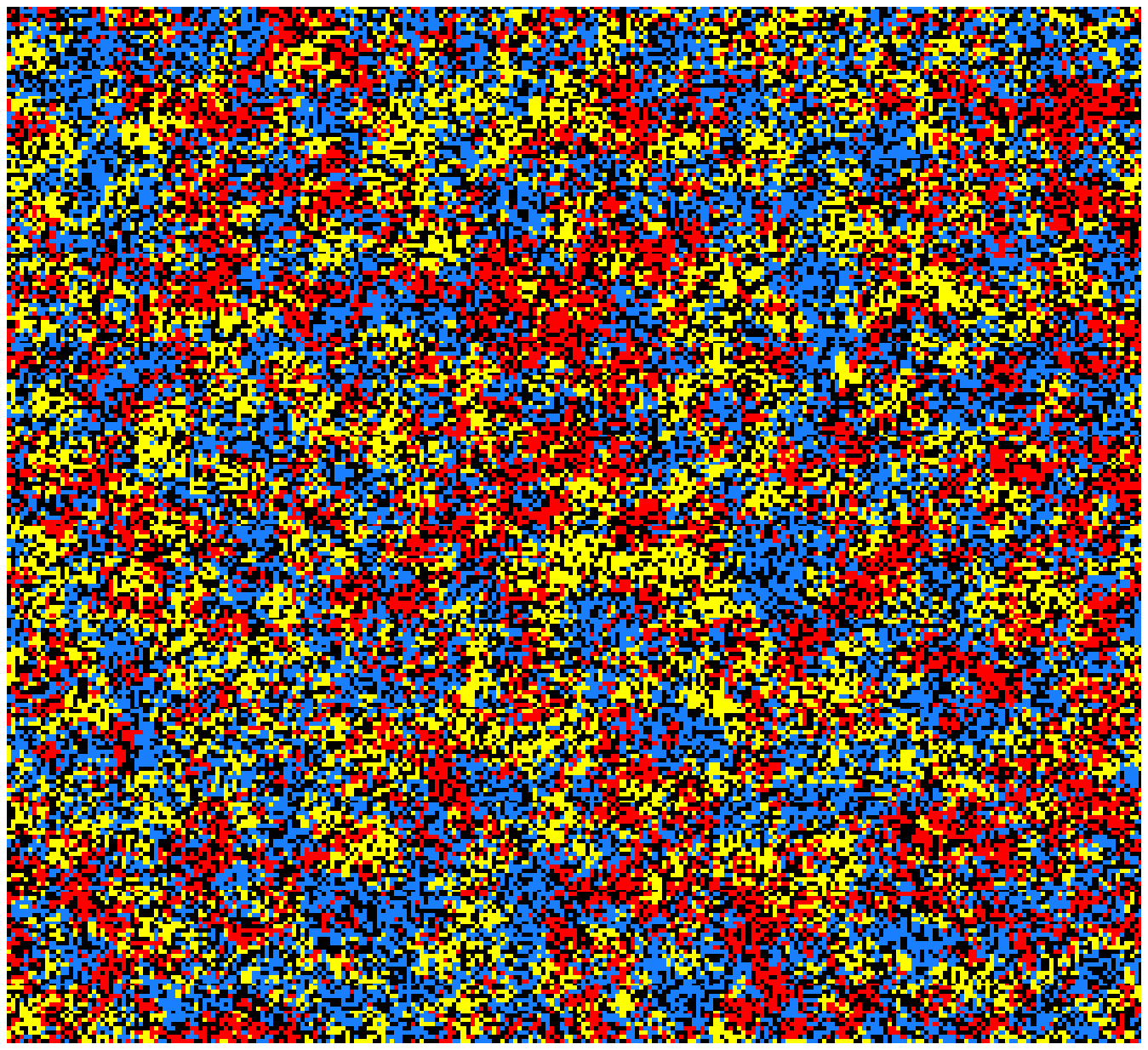}} \
\subfloat[]{\label{fig1:snpsh3}
\includegraphics[width=0.145\textwidth]{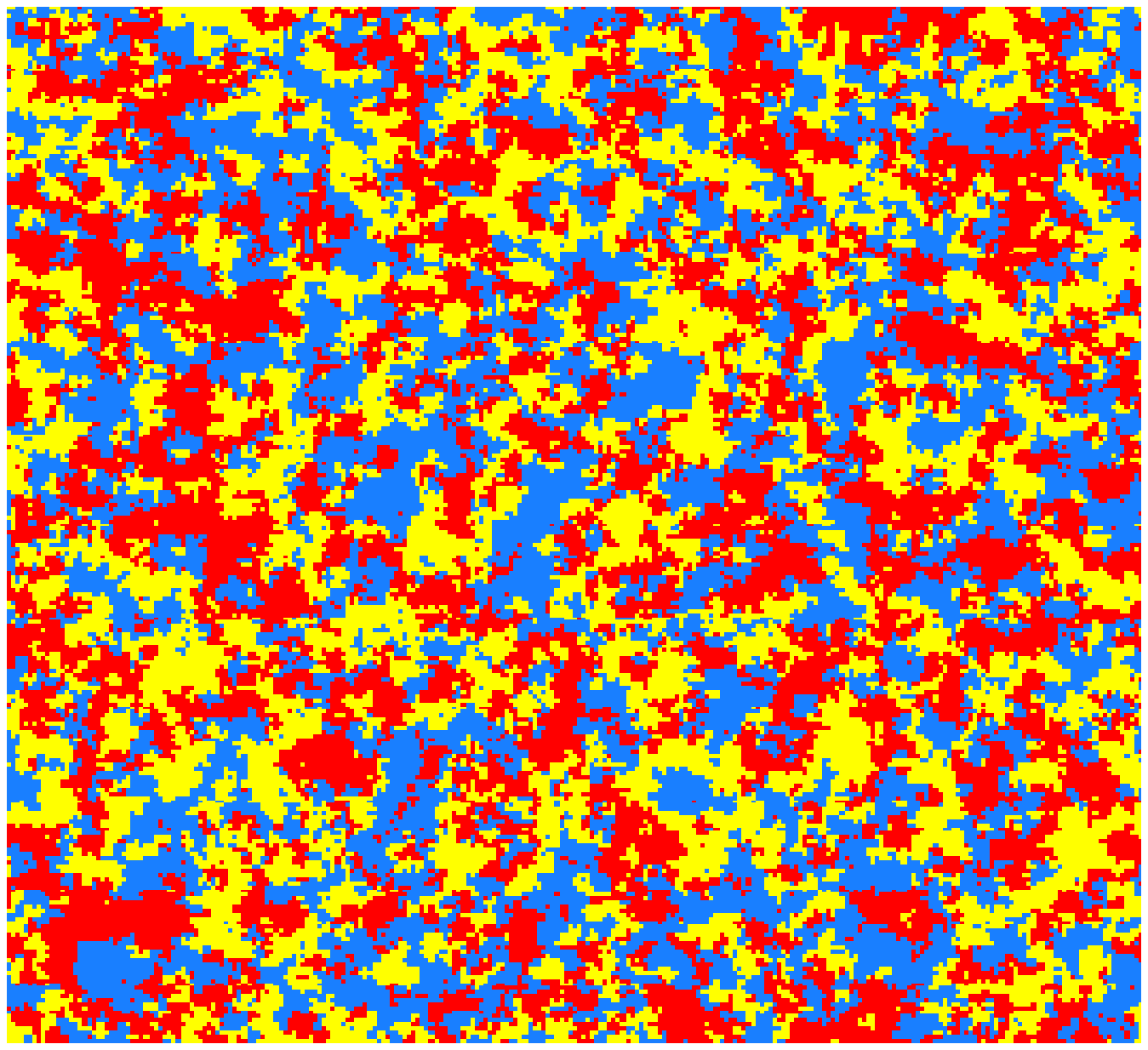}}
\end{center}
\caption{{\it (Color online.)} (a) Temporal evolution for the population
  densities of species $A$ (red/solid line), $B$(green/dashed), and $C$
  (blue/dash-dotted) with symmetric reaction rates $k_a = k_b = k_c =0.5$ and 
  without site occupation restriction (model 1), with unequal initial 
  densities $a(0) = 2/3$, $b(0) = c(0) = 1/6$, averaged over 50 Monte Carlo 
  runs on a 256 $\times$ 256 square lattice.
  (b) Snapshot of the spatial particle distribution for a single simulation
  run at $t = 50$, and (c) at $t = 500$ MCS; (d): snapshot at $t = 500$ MCS, 
  for a system where at most one particle of either species is allowed per 
  site (model 2).
  (Color coding shows the majority species on each site; red/gray: species 
  $A$, yellow/light gray: $B$, blue/dark gray: $C$, black: empty site.)}
\end{figure}
We first report and discuss our Monte Carlo simulation results on a
256 $\times$ 256 square lattice with periodic boundary conditions.
The data are typically averaged over 50 Monte Carlo runs with different initial
configurations, where the particles of each species are placed randomly on the
lattice.
Figure~\ref{fig1:twodim} depicts the temporal evolution of the total population
densities in a system without site occupation number restrictions and with 
equal reaction rates $k_a = k_b = k_c = 0.5$ (labeled model 1 in 
Table~\ref{models}), but unequal initial densities
$a(0) = 2/3$, $b(0) = c(0) = 1/6$, along with two snapshots 
\ref{fig1:snpsh1},\ref{fig1:snpsh2} of their spatial distribution at different 
times.
Since the selection and reproduction processes are combined into a single step
in our model, the total population density $\rho$ is strictly conserved, and as
expected we therefore observe no spiral patterns that are characteristic of RPS
models without conservation law \cite{Matti}.
In the initial time regime, we see distinct decaying population oscillations in
Fig.~\ref{fig1:twodim}, and inhomogeneous species clusters in the snapshot
Fig.~\ref{fig1:snpsh1}.
As time progresses, the amplitude of the oscillating fluctuations decreases
quickly, and also the spatial distribution and species cluster size become more
stable and homogeneous (Fig.~\ref{fig1:snpsh2}).
Our (fairly large) system eventually settles in a coexistence state with small
density fluctuations (Fig.~\ref{fig1:twodim} inset).
For comparison, Fig.~\ref{fig1:snpsh3} shows a snapshot in a system with
identical reaction rates and asymmetric initial densities, but with all site 
occupation numbers restricted to at most a single particle (model 2); one 
observes the same small cluster structure as in the absence of occupation 
restrictions.

\begin{figure}[!t]
\includegraphics[width=0.48\textwidth]{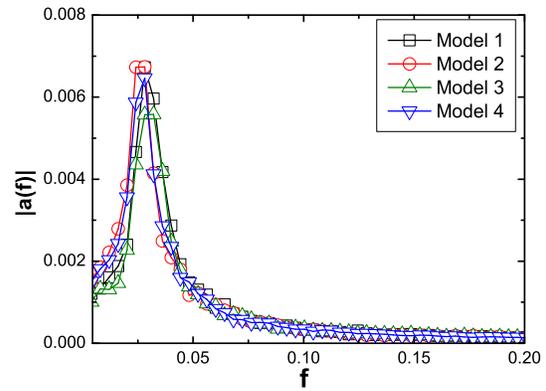}
\caption{{\it (Color online)}. \label{fig2:2dftsa} Signal Fourier transform
  $|a(f)|$ of species $A$ density data on a 256 $\times$ 256 square lattice
  with initial population densities $a(0) = b(0) = c(0) = 1/3$ for the four
  model variants described in Table~\ref{models}, averaged over 50 Monte Carlo
  simulation runs.}
\end{figure}
In Fig.~\ref{fig2:2dftsa} we show the absolute values of the Fourier
transformed population density signals $|a(f)|$, as obtained from averaging 50
Monte Carlo simulation runs for the four different model variants listed in
Table~\ref{models}, in this case with equal initial densities
$a(0) = b(0) = c(0) = 1/3$.
Recall that mean-field theory predicts a regular, undamped oscillation
frequency $f \approx 0.046$.
From the simulation data, we determine the characteristic peak frequency
$f \approx 0.028$, which evidently governs oscillatory fluctuations; however,
the finite width of the Fourier peak in Fig.~\ref{fig2:2dftsa} reflects that
the population oscillations are damped and will cease after a finite
characteristic relaxation time.

Moreover, we see that even if spatial disorder and/or site occupancy
restrictions are incorporated in the model, the Fourier-transformed density
signals display practically the same frequency distribution and significant
peak locations.
Indeed, we find that in our simulations for model versions 1 and 3 with total
density $1$, the typical occupation number at each site remains $n \leq 2$
throughout the runs, which explains why the exclusion constraints in model
variants 2 and 4 do not have a large effect.
Thus, neither spatial disorder nor site occupancy restrictions change the
temporal evolution pattern of the system markedly.
This is in stark contrast with results for the two-species stochastic lattice
Lotka--Volterra model, for which one finds (i) very pronounced spatio-temporal
structures in the species coexistence regime \cite{Ivan}; (ii) large
fluctuations that strongly renormalize the characteristic population
oscillation frequency \cite{Ivan,Mark}; (iii) an extinction threshold for the
predator species induced by local density restrictions on the prey \cite{Ivan};
and (iv) considerable enhancement of the asymptotic densities of both species
caused by spatial variability of the predation rate \cite{Ulrich}.

\begin{figure}[!t]
\subfloat[]{\label{fig3:2coraa}
\includegraphics[width=0.24\textwidth]{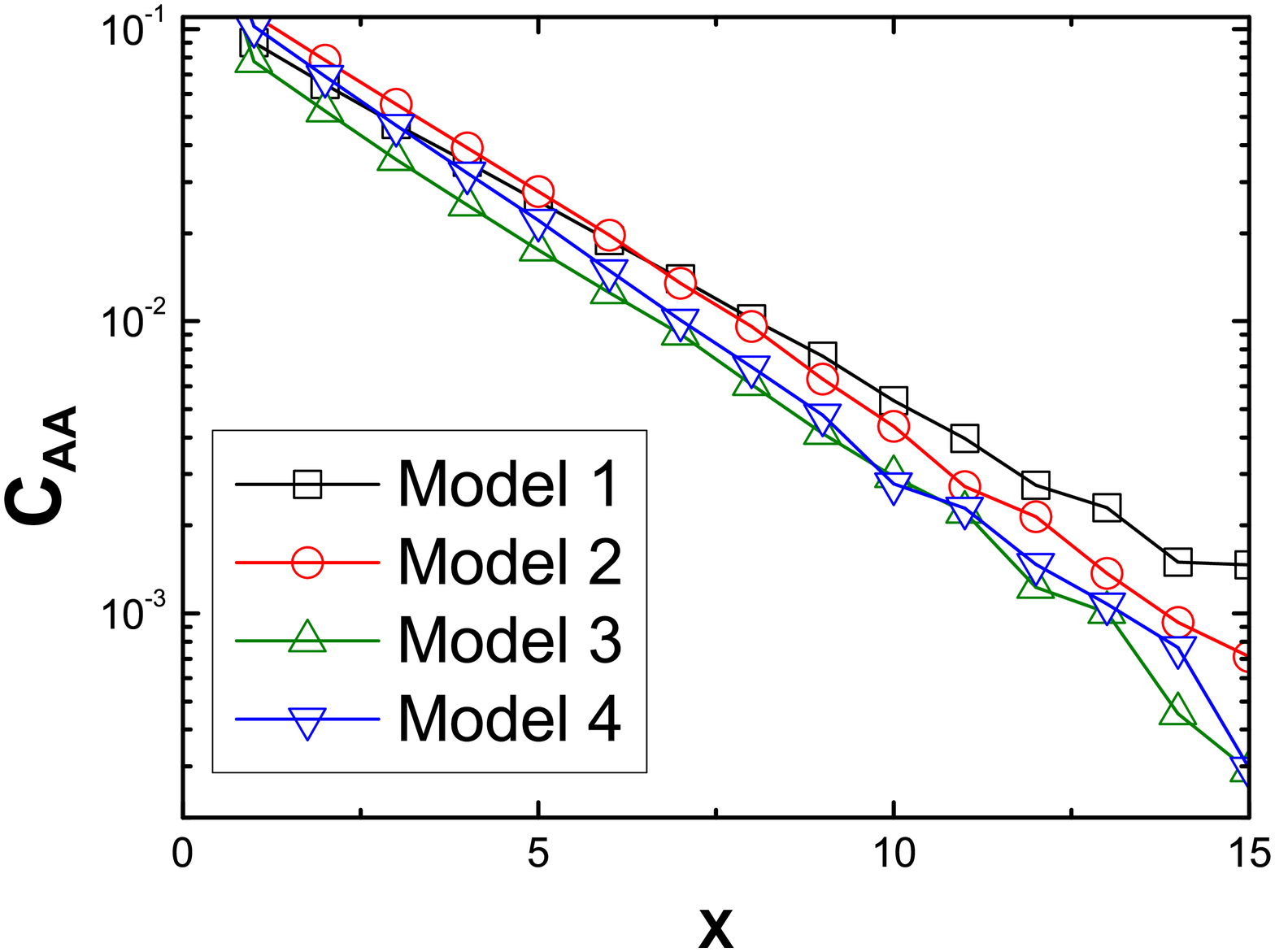}} \hskip -0.5cm
\subfloat[]{\label{fig3:2corab}
\includegraphics[width=0.252\textwidth]{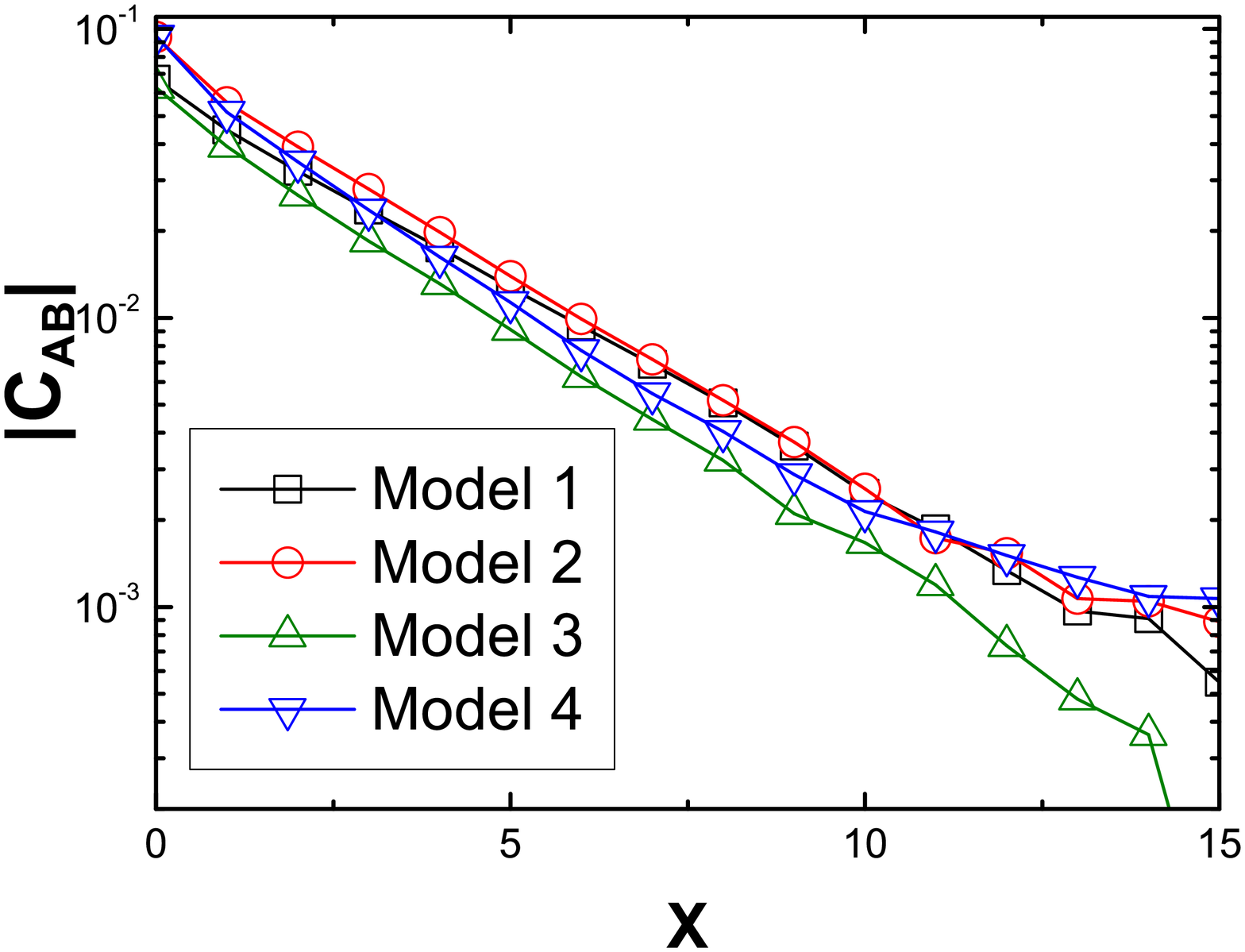}}
\caption{{\it (Color online.)} (a) Static density autocorrelation function
  $C_{AA}(x)$ and (b) cross-correlation function $C_{AB}(x)$ 
  (linear-$log_{10}$ plots) measured at $t = 250$ MCS for the four model 
  variants described in Table~\ref{models}, with initial population densities 
  $a(0) = b(0) = c(0) = 1/3$.}
\end{figure}
\begin{table}[!b]
\begin{tabular}{|l|l|l|l|l|} \hline
  & Model 1 & Model 2 & Model 3 & Model 4 \\ \hline
  $l_{AA}$ & 3.27 $\pm$ 0.02 & 2.92 $\pm$ 0.02 & 2.64 $\pm$ 0.03 &
  2.59 $\pm$ 0.01 \\ \hline
  $l_{AB}$ & 2.86 $\pm$ 0.08 & 2.35 $\pm$ 0.09 & 2.40 $\pm$ 0.06 &
  1.99 $\pm$ 0.09 \\ \hline
\end{tabular}
\caption{\label{sycorl} Correlation lengths $l_{AA}$ for the autocorrelation
  function and $l_{AB}$ for the cross-correlation function (in units of the
  lattice spacing) obtained for the four model variants of Table~\ref{models}
  with symmetric reaction rates.}
\end{table}
In order to study the effect of spatial disorder and site occupation 
restriction on emerging correlations in our stochastic RPS models, we have 
determined the equal-time two-point correlation functions in the 
quasi-stationary (long-lived) coexistence state illustrated for models 1 and 2
in Figs.~\ref{fig1:snpsh2} and \ref{fig1:snpsh3}, respectively.
These static correlation functions can quantitatively capture the emerging
spatial structures in the lattice.
Figures~\ref{fig3:2coraa} and \ref{fig3:2corab} depict the autocorrelation
function $C_{AA}(x)$ and the cross-correlation function $C_{AB}(x)$ as obtained
for our four models (see Table~\ref{models}), which all are seen to decay
exponentially with distance, i.e., $C_{AA}(x)\propto e^{-|x| /l_{AA}}$ and 
$C_{AB}(x)\propto e^{-|x|/l_{AB}}$.
From these log-normal plots, we have extracted the associated correlation
length $l_{AA}$ and typical species separation distance $l_{AB}$; the results
are listed in Table~\ref{sycorl}. 
It is worth noticing that in systems exhibiting spiralling patterns, as in the 
four-state RPS model without conservation law, the correlation functions 
$C_{AA}(x)$ and $C_{AB}(x)$ do not fall off exponentially but exhibit (damped) 
oscillations, see e.g. Refs.~\cite{Reichenbach3,Reichenbach4}.
Site occupation restrictions clearly have the effect of reducing both
correlation lengths.
Also, as is the case for the two-species lattice Lotka--Volterra system
\cite{Ulrich}, rendering the reaction rate a quenched random variable for each
site leads to more localized population and activity patches, characterized by
markedly smaller correlation and typical separation lengths.

\begin{figure}[!t]
\includegraphics[width=0.40\textwidth]{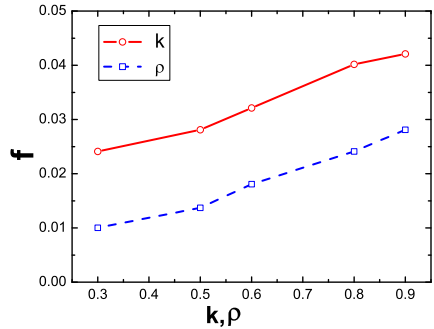}
\caption{{\it (Color online.)} \label{fig4:ratden} Variation of the 
  characteristic peak frequency in the density Fourier signal $|a(f)|$ with 
  the total density $\rho$ and homogeneous, symmetric reaction rate $k$, for 
  RPS simulations on a 256 $\times$ 256 square lattice with equal initial
  densities, run for 1000 MCS.}
\end{figure}
The influence of varying (homogeneous and symmetric) reaction rates and
modifying the total (conserved) population density is explored in
Fig.~\ref{fig4:ratden}, which shows the dependence of the characteristic
Fourier peak frequency $f$ on $k$ and $\rho$.
We find that $f$ scales roughly linearly with both the total density $\rho$ and
the reaction rate $k$, in accord with the mean-field prediction
$f \propto \rho \, k$,  see Sec.~\ref{meanft}.
We have also checked that switching off nearest-neighbor hopping
(setting $D = 0$), thus allowing particle spreading only via the nonlinear
reaction processes (\ref{react}), essentially leaves the stochastic RPS
system's features intact.

\begin{figure}[!b]
\subfloat[]{\label{fig5:exttwo}
\includegraphics[width=0.25\textwidth]{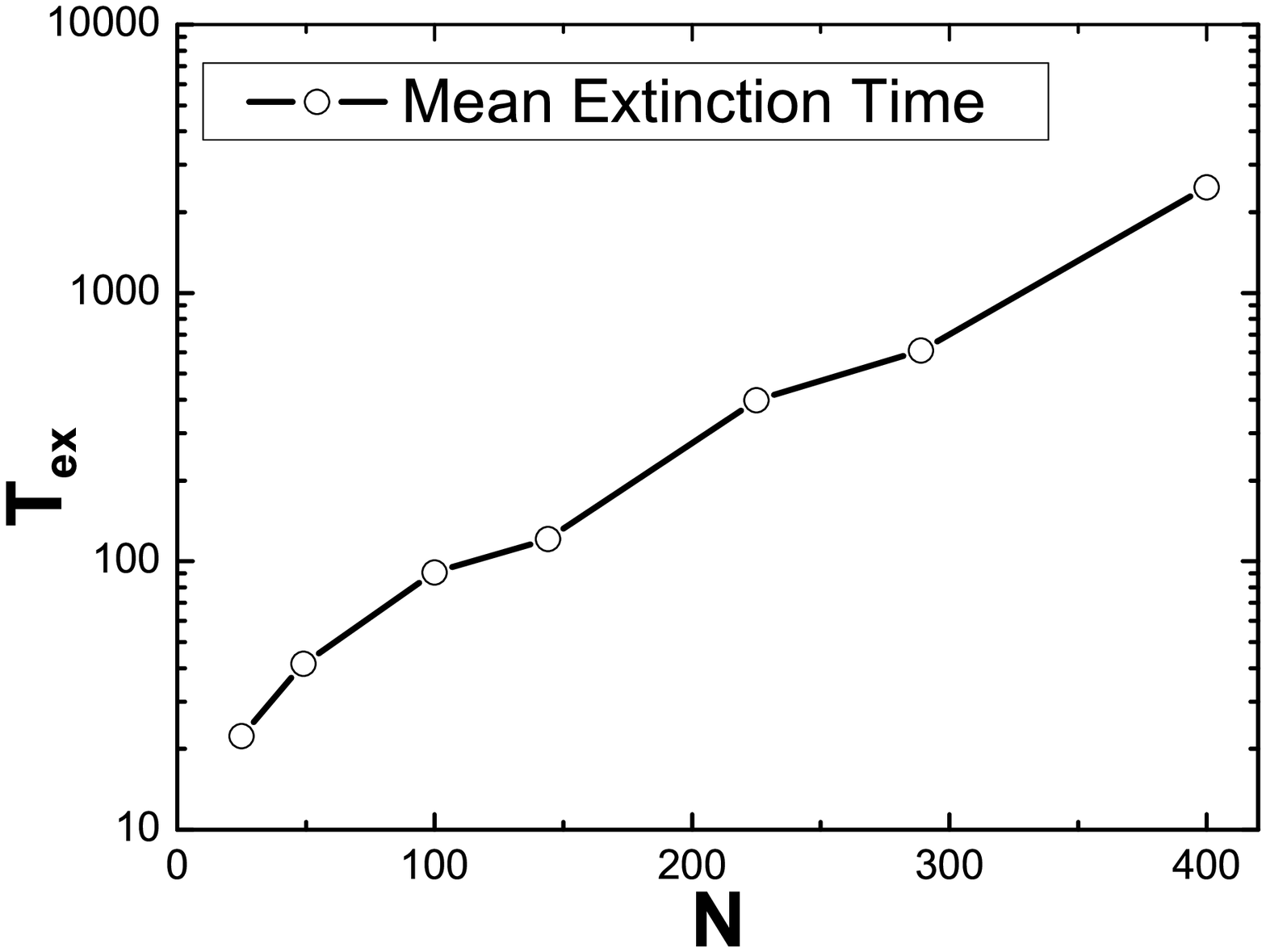}} \hskip -0.7cm
\subfloat[]{\label{fig5:exdis2}
\includegraphics[width=0.25\textwidth]{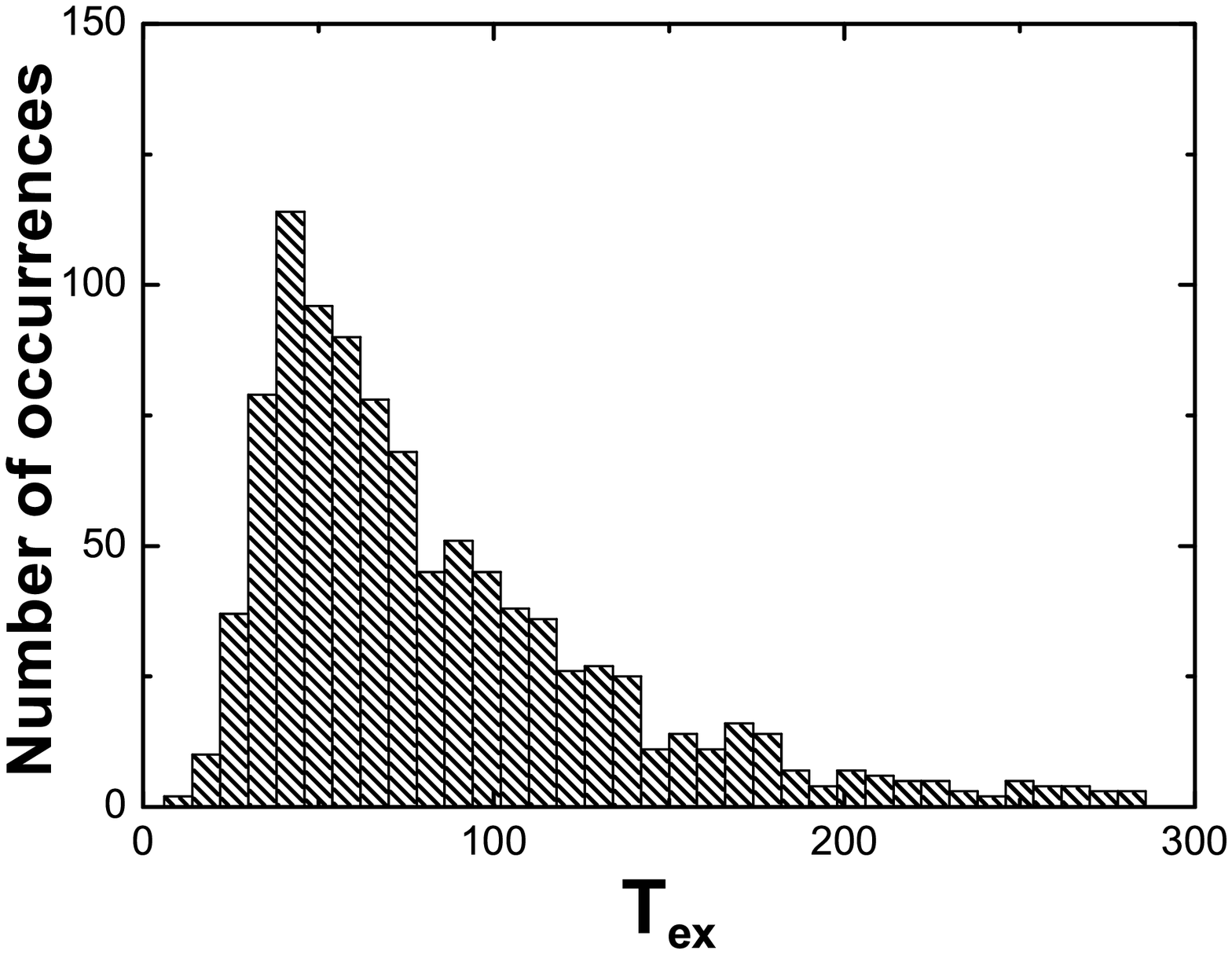}}
\caption{(a) Mean extinction time as function of lattice size $N$
  (linear-$log_{10}$ plot), obtained from averages over 50 Monte Carlo runs, 
  for small two-dimensional lattice RPS systems in the absence of site 
  restrictions and with symmetric reaction rates $k_a = k_b = k_c = 0.5$ 
  (model 1), and equal initial population densities $a(0) = b(0) = c(0) = 1/3$.
  The data are for lattices with $N = 5 \times 5$, $7 \times 7$, 
  $10 \times 10$, $12 \times 12$, $15 \times 15$, $17 \times 17$, and
  $20 \times 20$ sites.
  (b) Histogram of measured extinction times for $N = 100$ sites.}
\end{figure}
Finally, we have also studied the mean extinction time as function of lattice
size $N$ for small two-dimensional stochastic lattice RPS systems, here of the
model 1 variety with homogeneous symmetric reaction rates
$k_a = k_b = k_c = 0.5$ and equal initial densities $a(0) = b(0) = c(0) = 1/3$.
We recall that in any finite system displaying an absorbing stationary state,
stochastic fluctuations will eventually reach this absorbing configuration.
In the stochastic RPS model, one therefore expects two species to eventually
become extinct; however, reaching this absorbing state may take an enormous
amount of time, and will thus become practially unobservable on large lattices.
In fact, in two and higher dimensions one expects the mean extinction time
$T_{\rm ex}$ (here measured for the first species to die out) to scale
exponentially with system size $N$, since random fluctuations effectively have
to overcome a finite barrier in order to follow an `optimal' path towards
extinction.
As depicted in Fig.~\ref{fig5:exttwo}, we indeed observe
$\ln T_{\rm ex}(N) \sim N$, consistent with the prediction on the coexistence
state stability reported in Refs.~\cite{Reichenbach1,Reichenbach4}.
The associated distributions of extinction times are described by neither
Poisson nor Gaussian distributions (e.g., the means are considerably larger
than the most likely values), but display long `fat' tails at large extinction
times, see Fig.~\ref{fig5:exdis2}.
We expect similar features in model variant 2, in accord with the remarkably
long-live species coexistence observed in Ref.~\cite{Tainaka1}.

\subsection{Two-dimensional stochastic RPS system: asymmetric rates}
\label{twodas}

\begin{table}[!b]
\begin{tabular}{|c|c|c|} \hline
  \textbf{Model} & \textbf{Reaction rates} & \textbf{Site restriction} \\
  \hline
  \textbf{1} & $k_a = 0.2, k_b = 0.5, k_c = 0.8$ & no restriction \\ \hline
  \textbf{2} & $k_a = 0.2, k_b = 0.5, k_c = 0.8$ & at most one \\ \hline
  \textbf{3} & $k_a \in [0,0.4]$, $k_b=0.5$, $k_c=0.8$ & no restriction \\
  \hline
  \textbf{4} & $k_a \in [0,0.4]$, $k_b=0.5$, $k_c=0.8$ & at most one \\
  \hline
\end{tabular}
\caption{\label{models-as} List of stochastic lattice RPS model variants with
  asymmetric rates.
  While $k_b = 0.5$ and $k_c = 0.8$ are held fixed in all four variants, we
  set $k_a = 0.2$ in models 1 and 2, whereas we took $k_a$ to be a random
  variable uniformly distributed in $[0,0.4]$ in models 3 and 4.}
\end{table}
\begin{figure}[!t]
\subfloat[]{\label{fig6:astwod}
\includegraphics[width=0.45\textwidth]{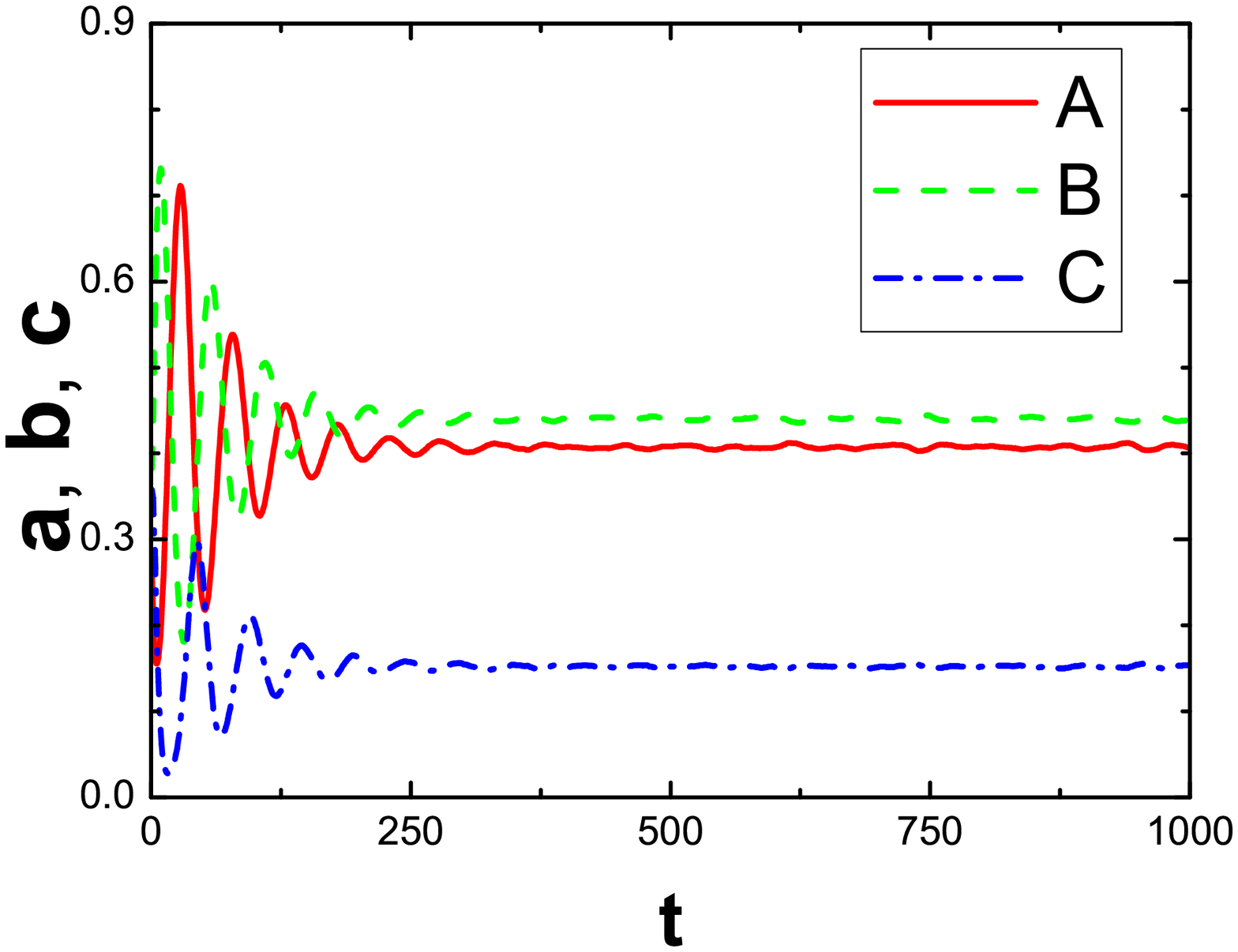}}
\smallskip \\
\subfloat[]{\label{fig6:snpsh1}
\includegraphics[width=0.145\textwidth]{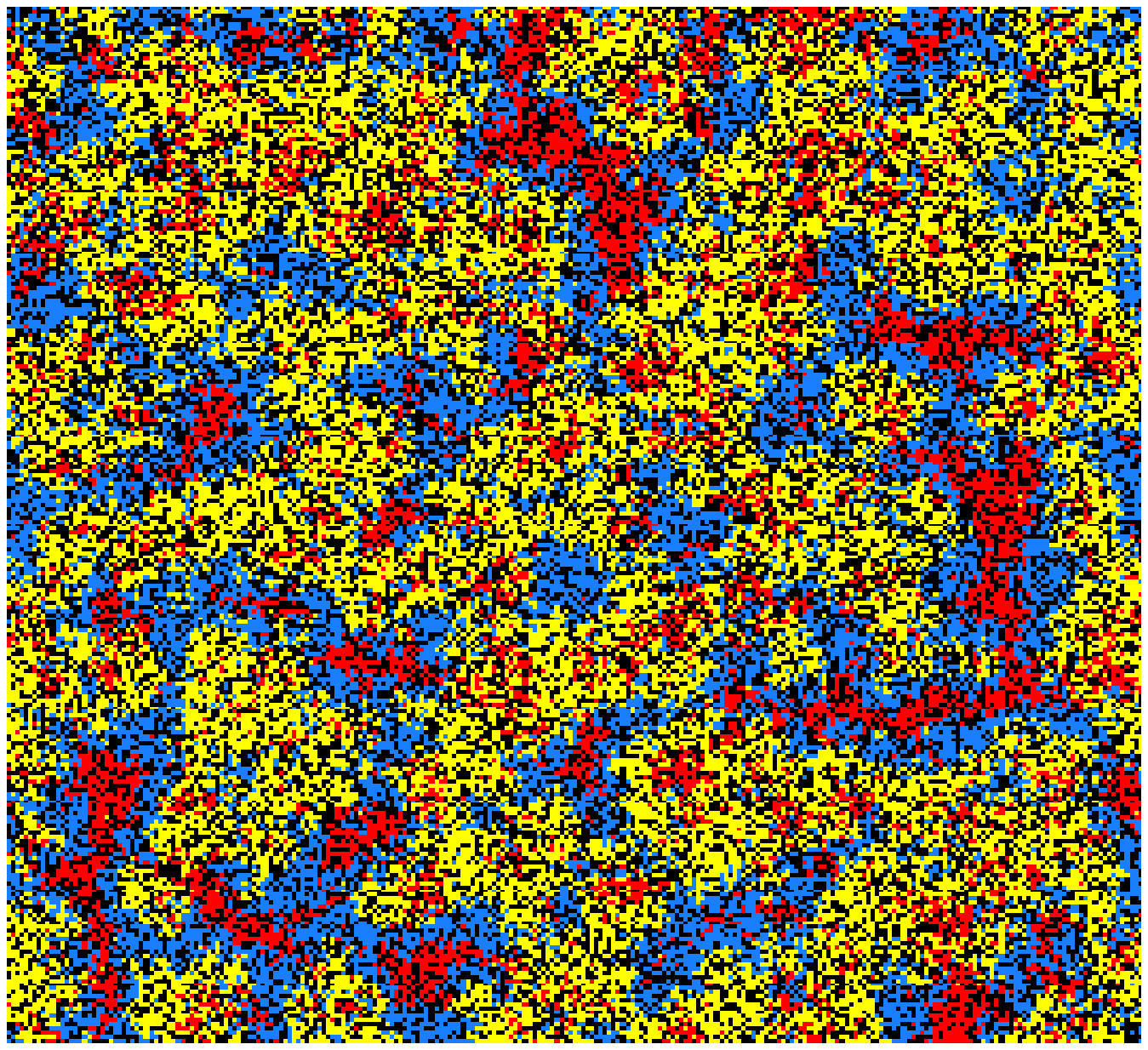}} \
\subfloat[]{\label{fig6:snpsh2}
\includegraphics[width=0.145\textwidth]{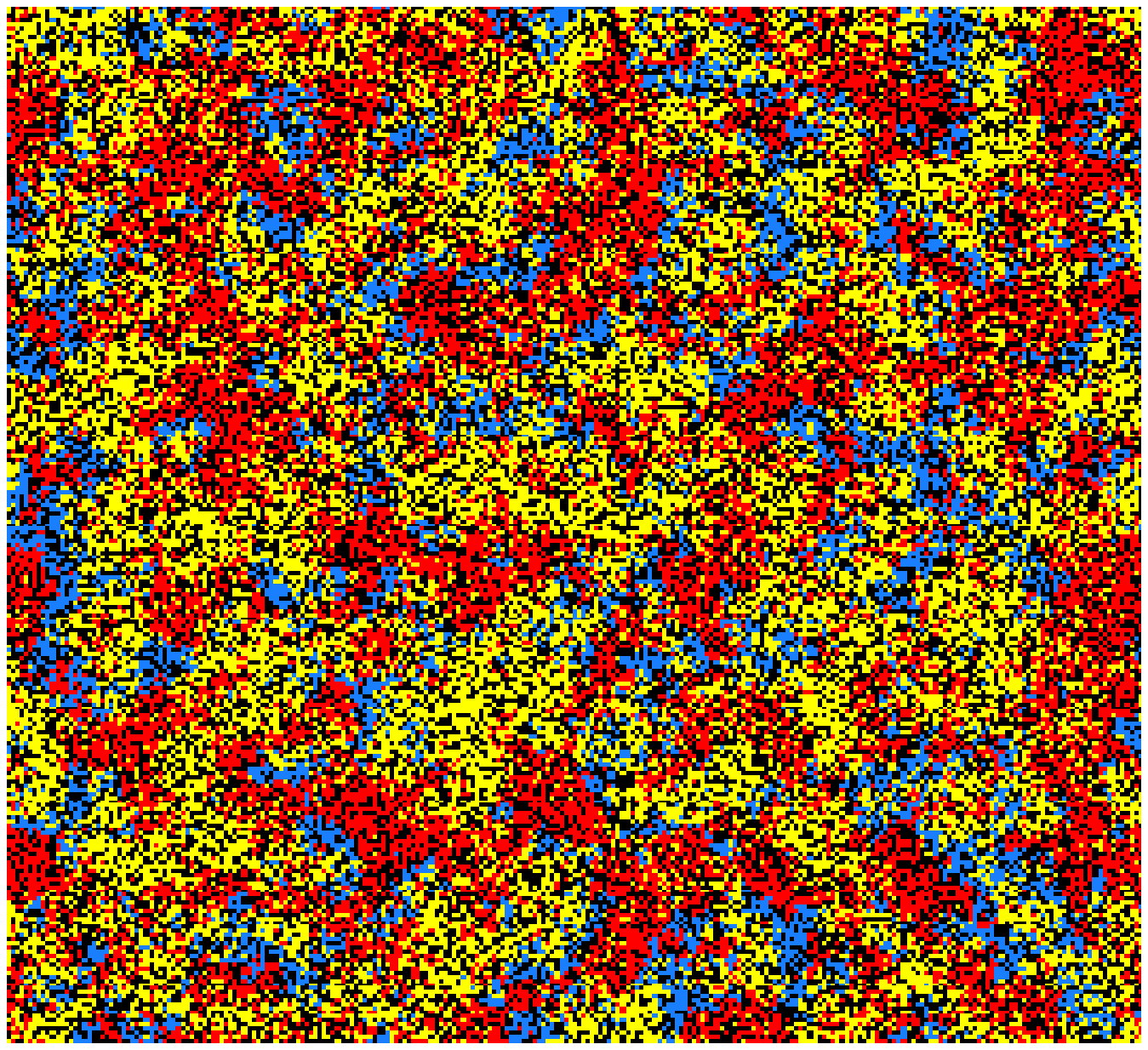}} \
\subfloat[]{\label{fig6:snpsh3}
\includegraphics[width=0.145\textwidth]{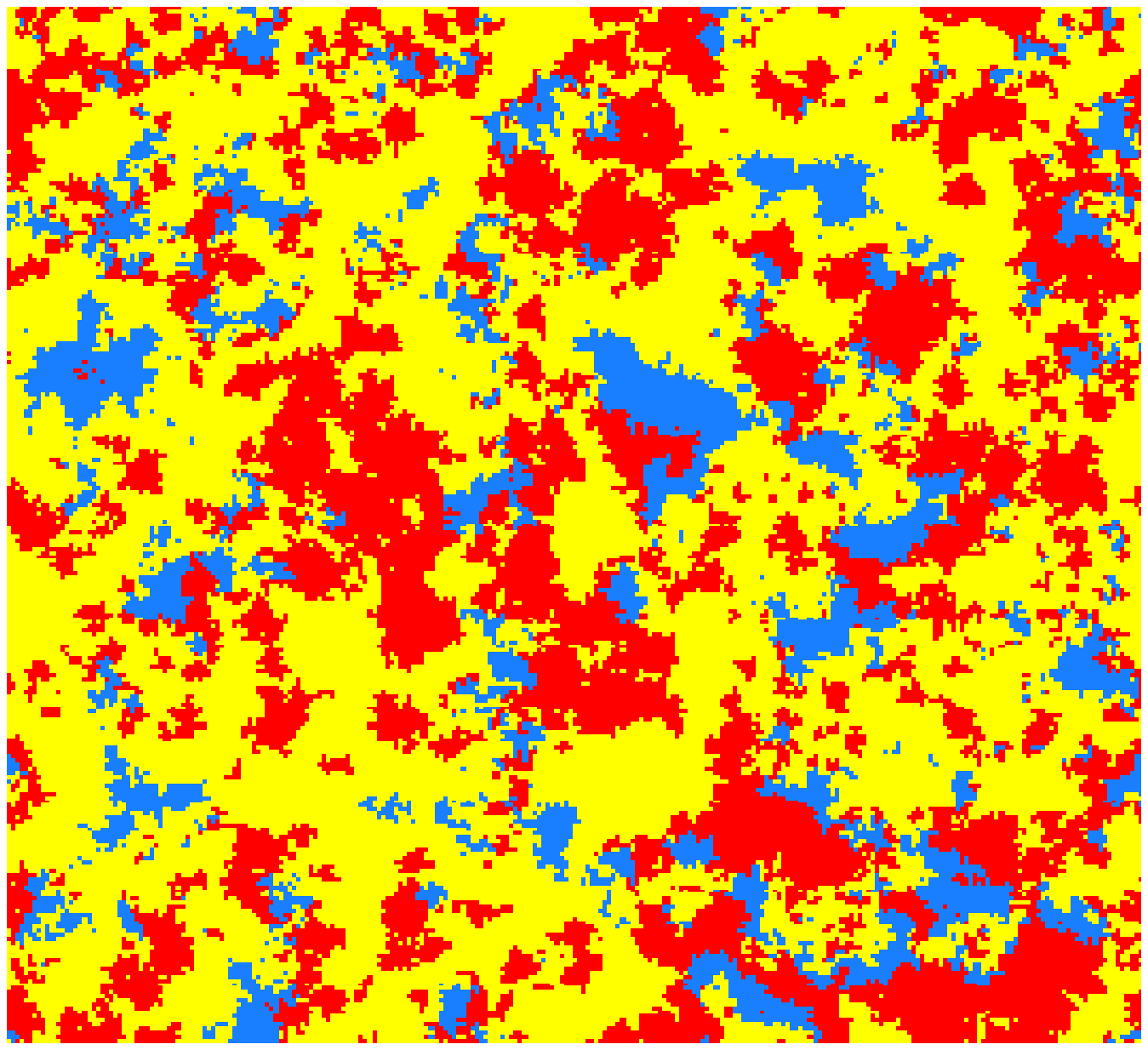}}
\caption{{\it (Color online.)} (a) Temporal evolution for the population
  densities of species $A$ (red/solid line), $B$ (green/dashed), and $C$
  (blue/dash-dotted) with asymmetric reaction rates $k_a = 0.2$, $k_b = 0.5$,
  $k_c = 0.8$ and without site occupation restriction (model 1), with equal 
  initial densities $a(0) = b(0) = c(0) = 1/3$, averaged over 50 runs on a 
  256 $\times$ 256 square lattice.
  (b) Snapshot of the spatial particle distribution in a single simulation run
  at $t = 50$, and (c) at $t = 500$ MCS; (d): snapshot at $t = 500$ MCS, for a
  system where at most one particle of either species is allowed per site
  (model 2).
  (Majority species coloring: red/gray: $A$, yellow/light gray: $B$, blue/dark 
  gray: $C$, black: empty.)}
\end{figure}
Next we turn to a stochastic RPS system with asymmetric reaction rates and
consider the various model variants specified in Table~\ref{models-as}
together with the reactions (\ref{react}).
Figure~\ref{fig6:astwod} shows the time evolution for the three species'
densities in a system with constant rates $k_a = 0.2$, $k_b = 0.5$, and
$k_c = 0.8$.
From our simulations for model version 1, we infer the asymptotic population
densities (with statistical errors)
$(0.40 \pm 0.01, 0.45 \pm 0.01, 0.15 \pm 0.01)$, which follow the trends of the
mean-field results $(a^*,b^*,c^*) = (0.33,0.53,0.13)$.
As becomes apparent in the snapshots \ref{fig6:snpsh1} and \ref{fig6:snpsh2} 
for model variant 1 without site restrictions, and \ref{fig6:snpsh3} for a
system with at most a single particle per site (model 2), particles of the same
species form distinctive spatial clusters.
The effect of the reaction rate asymmetry on the equal-time auto- and
cross-correlation functions is shown in Figs.~\ref{fig7:2aucaa} and
\ref{fig7:2aucab}, respectively, with the ensuing correlation lengths and
typical separation distances listed in Table~\ref{ascorl}.
Note that the autocorrelation length $l_{CC}$ for species $C$ is smaller than
$l_{AA}$, and $l_{BB}$, which is largest.
This is consistent with the long-time densities in the (quasi-stationary)
coexistence state, given our observation that the overall particle density is
roughly uniform.
\begin{figure}[!b]
\subfloat[]{\label{fig7:2aucaa}
\includegraphics[width=0.24\textwidth]{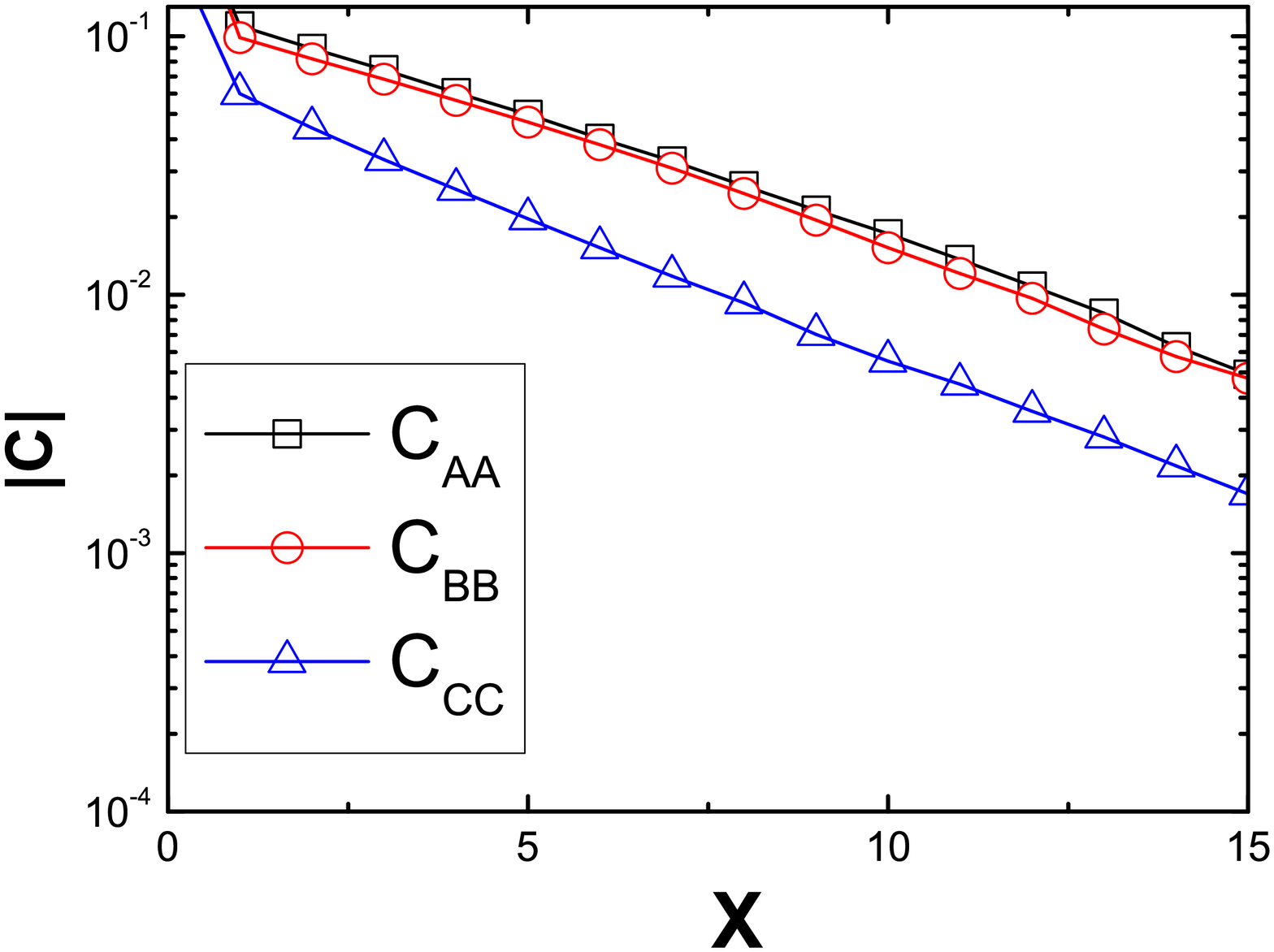}} \hskip -0.5cm
\subfloat[]{\label{fig7:2aucab}
\includegraphics[width=0.24\textwidth]{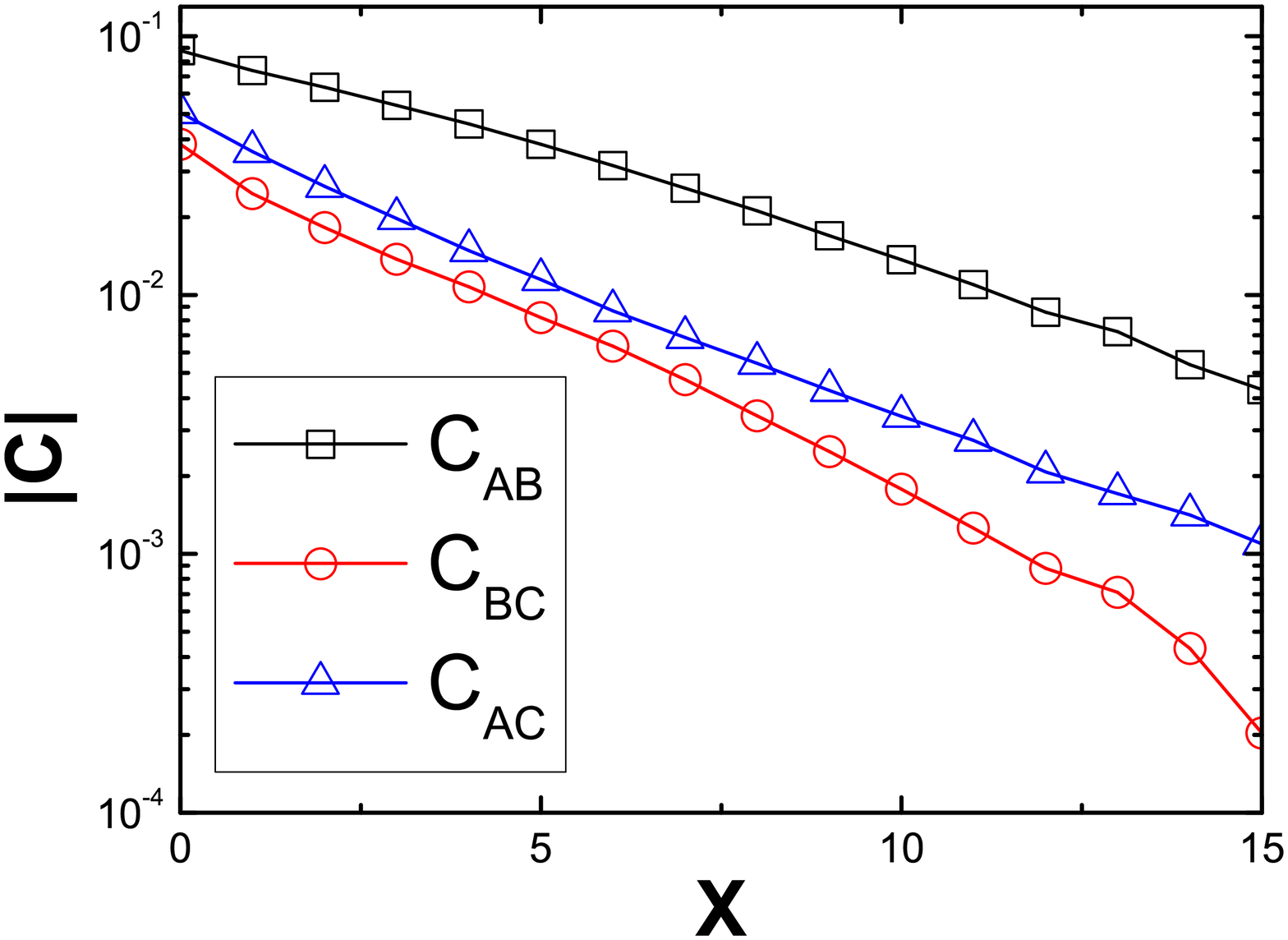}}
\caption{{\it (Color online.)} (a) Equal-time autocorrelation functions
  $C_{AA}(x)$, $C_{BB}(x)$, $C_{CC}(x)$ at $t = 1000$ MCS for the model
  described in Fig.~6.
  (b) Equal-time cross-correlation functions $C_{AB}(x)$, $C_{BC}(x)$,
  $C_{AC}(x)$ (linear-$log_{10}$ plots).}
\end{figure}
\begin{table}[!t]
\begin{tabular}{|c|c|c|} \hline
  $l_{AA}$ & $l_{BB}$ & $l_{CC}$ \\ \hline
  5.24 $\pm$ 0.03 & 5.67 $\pm$ 0.08 & 3.46$\pm $0.05 \\ \hline
  $l_{AB}$ & $l_{BC}$ & $l_{AC}$ \\ \hline
  6.68 $\pm$ 0.20 & 3.68 $\pm$ 0.07 & 3.33 $\pm$ 0.05 \\ \hline
\end{tabular}
\caption{\label{ascorl} Correlation lengths (top) inferred from the
  autocorrelation functions and typical separation distances (bottom) obtained
  from the cross-correlation functions (in units of the lattice spacing)
  measured for the RPS model with asymmetric but homogeneous reaction rates
  $k_a = 0.2$, $k_b = 0.5$, and $k_c = 0.8$.}
\end{table}

\begin{figure}[!t]
\includegraphics[width=0.44\textwidth]{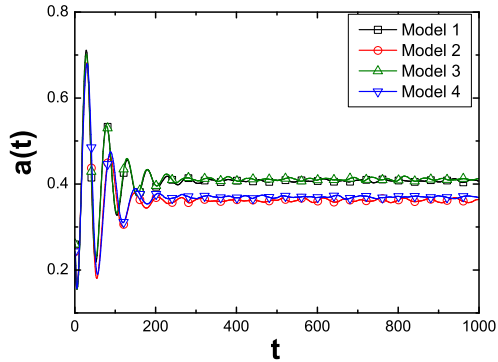}
\caption{{\it (Color online.)} \label{fig8:asdisd} Time evolution for the
  population density $a(t)$ of species $A$ for four model variants with
  asymmetric reaction rates, namely with $k_b = 0.5$, $k_c = 0.8$ and either
  uniformly $k_a = 0.2$, or drawn from a flat distribution $[0, 0.4]$, with and
  without site restrictions (see the listing in Table~\ref{models-as}).
  The initial densities are $a(0) = b(0) = c(0) = 1/3$, and the data stem from
  averages over 50 runs on a 256 $\times$ 256 square lattice.}
\end{figure}
\begin{figure}[!t]
\includegraphics[width=0.43\textwidth]{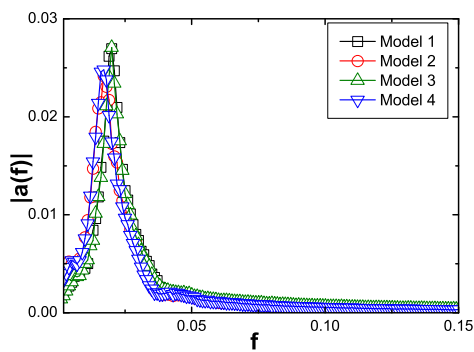}
\caption{{\it (Color online.)} \label{fig9:asdisf} Signal Fourier transform
  $|a(f)|$ for the four model variants described in Table~\ref{models-as}.
  The characteristic frequency comes out to be $f \approx 0.021$.}
\end{figure}
\begin{figure}[!t]
\subfloat[]{\label{fig10:2artcaa}
\includegraphics[width=0.24\textwidth]{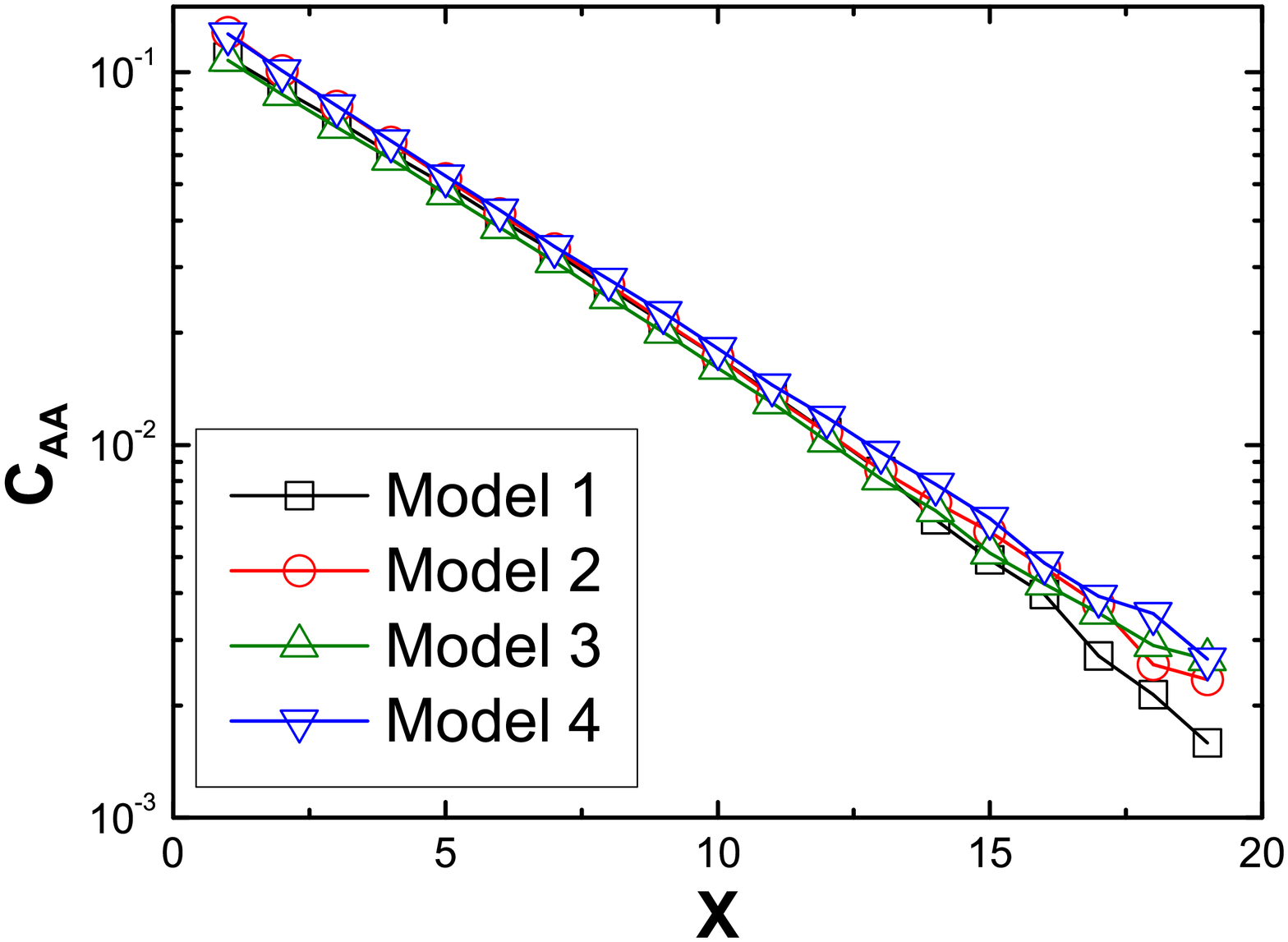}}
\hskip -0.5cm \subfloat[]{\label{fig10:2artcab}
\includegraphics[width=0.24\textwidth]{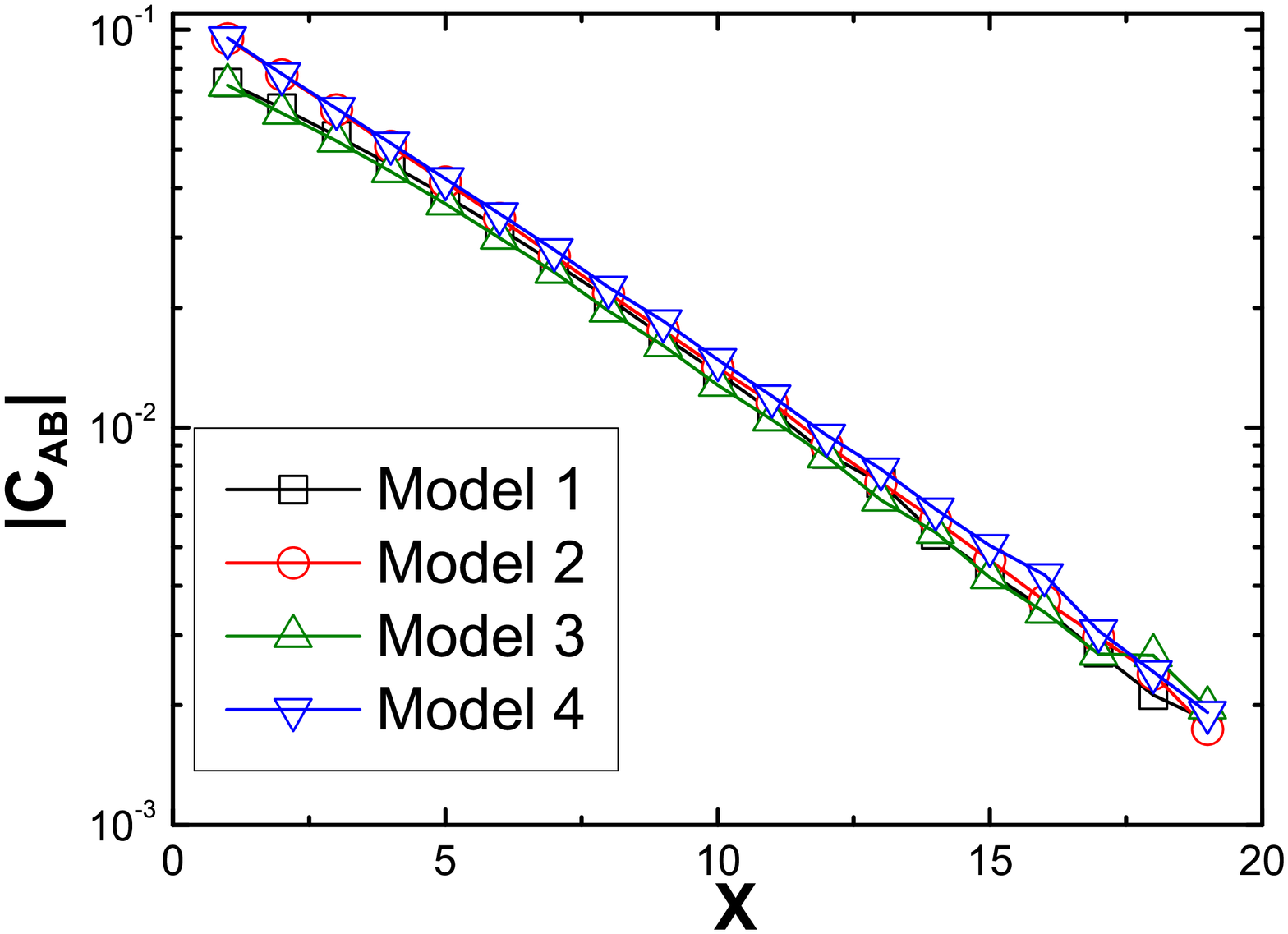}}
\caption{{\it (Color online.)} (a) Equal-time autocorrelation function
  $C_{AA}(x)$ and (b) cross-correlation functions $C_{AB}(x)$ 
  (linear-$log_{10}$ plots) at $t = 1000$ MCS for the four model variants 
  described in Table~\ref{models-as}.}
\end{figure}
As a last model variation, we allow the reaction rate $k_a$ to be a quenched
spatial random variable drawn from the flat distribution $[0, 0.4]$, such that
its average is still $0.2$, but hold $k_b = 0.5$ and $k_c = 0.8$ fixed.
Fig.~\ref{fig8:asdisd} compares the time evolution for these disordered systems
with and without site restrictions with the corresponding homogeneous models.
Once again, we see that spatial variability in the reaction rate even in this
asymmetric setting has very little effect.
As can be seen from the Fourier signal peak in Fig.~\ref{fig9:asdisf}, the
characteristic frequency comes out to be $f \approx 0.021$ for all four
asymmetric model variants investigated here, and Figs.~\ref{fig10:2artcaa} and
\ref{fig10:2artcab} demonstrate that the disorder hardly modifies the spatial
decay of the auto- and cross-correlation functions either.

\subsection{One-dimensional Monte Carlo simulations}
\label{onedim}

\begin{figure}[!t]
\includegraphics[width=0.22\textwidth]{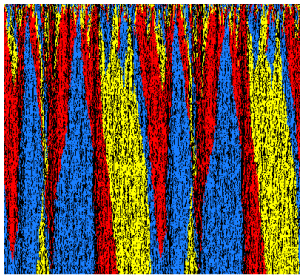}
\caption{{\it (Color online.)} \label{fig11:1dnorest} Time evolution (up to
  1000 Monte Carlo steps; from top to bottom) for a one-dimensional RPS model
  run with equal, homogeneous reaction rates $k_a = k_b = k_c = 0.5$, equal
  initial densities $a(0) = b(0) = c(0) = 1/3$, and in the absence of site
  occupancy restriction.
  (Only 10000 of the total 50000 lattice sites in this run are shown; majority
  species coloring in red/gray: $A$, yellow/light gray: $B$, blue/dark gray: 
  $C$, black: empty.)}
\end{figure}

\begin{figure}[!t]
\subfloat[]{\label{fig12:1drefl}
\includegraphics[width=0.22\textwidth]{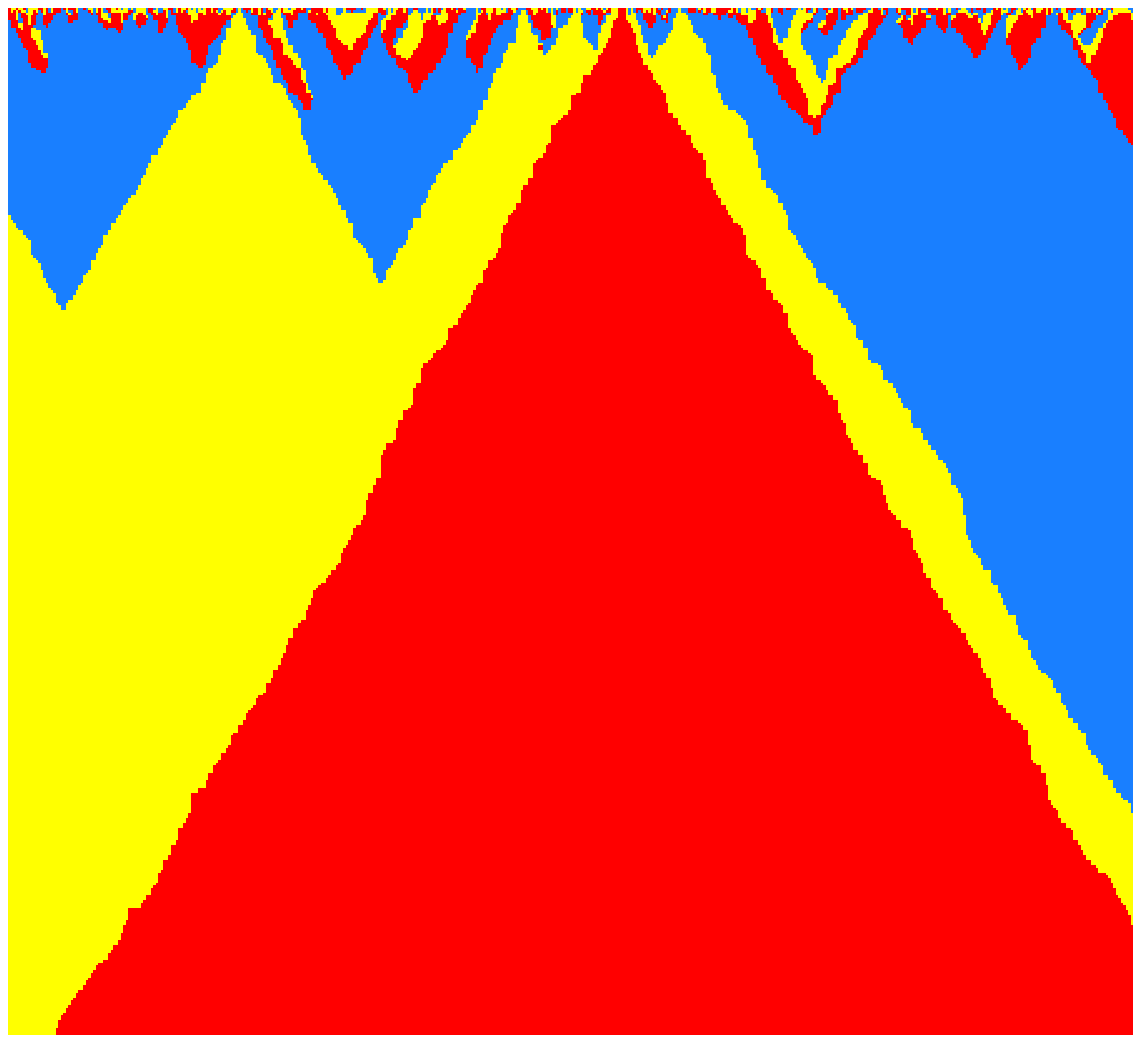}} \
\subfloat[]{\label{fig12:1dreem}
\includegraphics[width=0.22\textwidth]{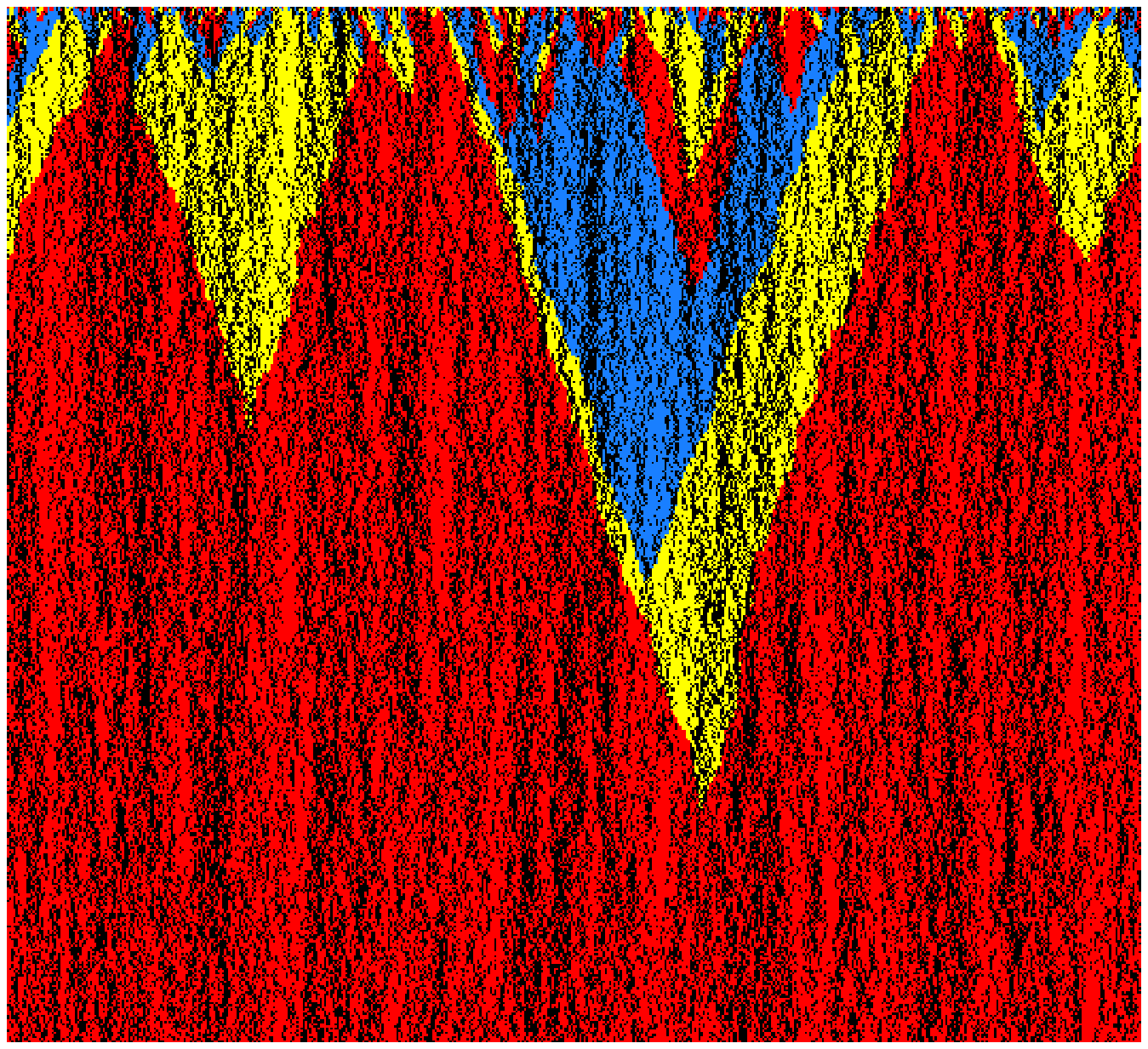}}
\caption{{\it (Color online.)} Time evolution (up to 1000 Monte Carlo steps;
  from top to bottom) for one-dimensional RPS model runs with equal,
  homogeneous reaction rates $k_a = k_b = k_c = 0.5$, equal initial densities
  (a) model 2: $a(0) = b(0) = c(0) = 1/3$, and (b) model 2': $a(0) = b(0) =
  c(0) = 0.2$ (model 2' and 4' refer to the corresponding model variants listed
  in Table~\ref{models} with total particle density less than $1$), where at
  most one particle of either species is allowed per site.
  (Only 10000 of the total 50000 lattice sites in these runs are shown;
  red/gray: $A$, yellow/light gray: $B$, blue/dark gray: $C$, black: empty.)}
\end{figure}

We have run simulations for all four model variants listed in
Table~\ref{models}, i.e., with/without site occupancy restriction; with/without
quenched spatial randomness in the reaction rates, in one dimension.
We find that only a single species ultimately survives and eventually occupies
the whole lattice no matter whether spatial disorder or site restrictions are
included in the model: as expected, the one-dimensional system will always
evolve towards one of the three extinction states where two of the three
species will die out.
While this phenomenon also occurs in two dimensions, in $d=1$ the species
coexist over a time that on average scales polynomially with the system size
(see below), i.e., extinction happens on a much shorter time scale than in two
dimensions (see Fig.~\ref{fig5:exttwo}).
Again, for equal (mean) reaction rates and initial densities, each species has
equal survival probability.
For comparison, the space-time plots of one-dimensional lattice simulations
without and with site occupancy restriction are depicted in
Figs.~\ref{fig11:1dnorest} and 12, respectively.
It is seen that individuals of identical species cluster together, and any
reactions are confined to the boundary separating the single-species domains.
When the occupancy of any site is restricted to a single particle of either
species, these domains form quickly and are very robust, even if not all sites
are filled, see Figs.~\ref{fig12:1drefl} and \ref{fig12:1dreem}.

\begin{figure}[!t]
\includegraphics[width=0.44\textwidth]{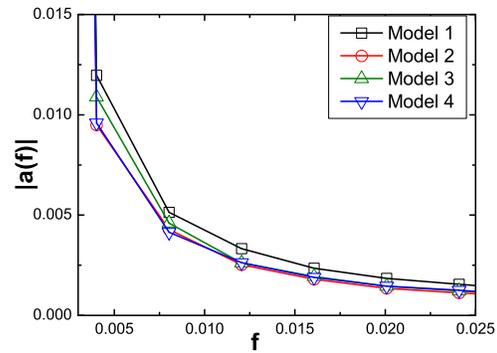}
\caption{{\it (Color online.)} \label{fig13:1dsift} Signal Fourier transform
  $|a(f)|$ for the four RPS model variants listed in Table~\ref{models}, in one
  dimension.
  The data is averaged over 50 Monte Carlo simulations on lattices with 50000
  sites.}
\end{figure}
\begin{figure}[!b]
\includegraphics[width=0.44\textwidth]{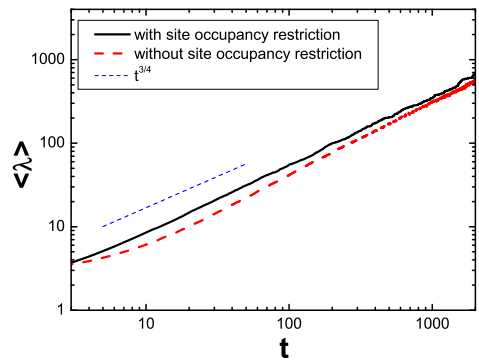}
\caption{{\it (Color online.)} \label{fig14:1ddoms} The time evolution 
  ($log_{10}-log_{10}$ plot) of the mean single-species domain size 
  $\langle \lambda(t) \rangle$ measured in one-dimensional Monte Carlo 
  simulation runs with 10000 lattice sites for RPS models with symmetric 
  reaction rates $k_a = k_b = k_c = 0.5$ and equal initial densities 
  $a(0) = b(0) = c(0) = 1/3$.
  The upper (black) curve shows the data for model 2 with site occupation
  restriction, see Fig.~\ref{fig12:1drefl}, whereas the lower (red/dashed) 
  graph pertains to model variant 1 without site occupancy restrictions, see
  Fig.~\ref{fig11:1dnorest}.
  For comparison, the blue/dotted straight line represents the predicted 
  $t^{3/4}$ power law.}
\end{figure}
\begin{figure}[!t]
\subfloat[]{\label{fig15:cmpaa}
\includegraphics[width=0.24\textwidth]{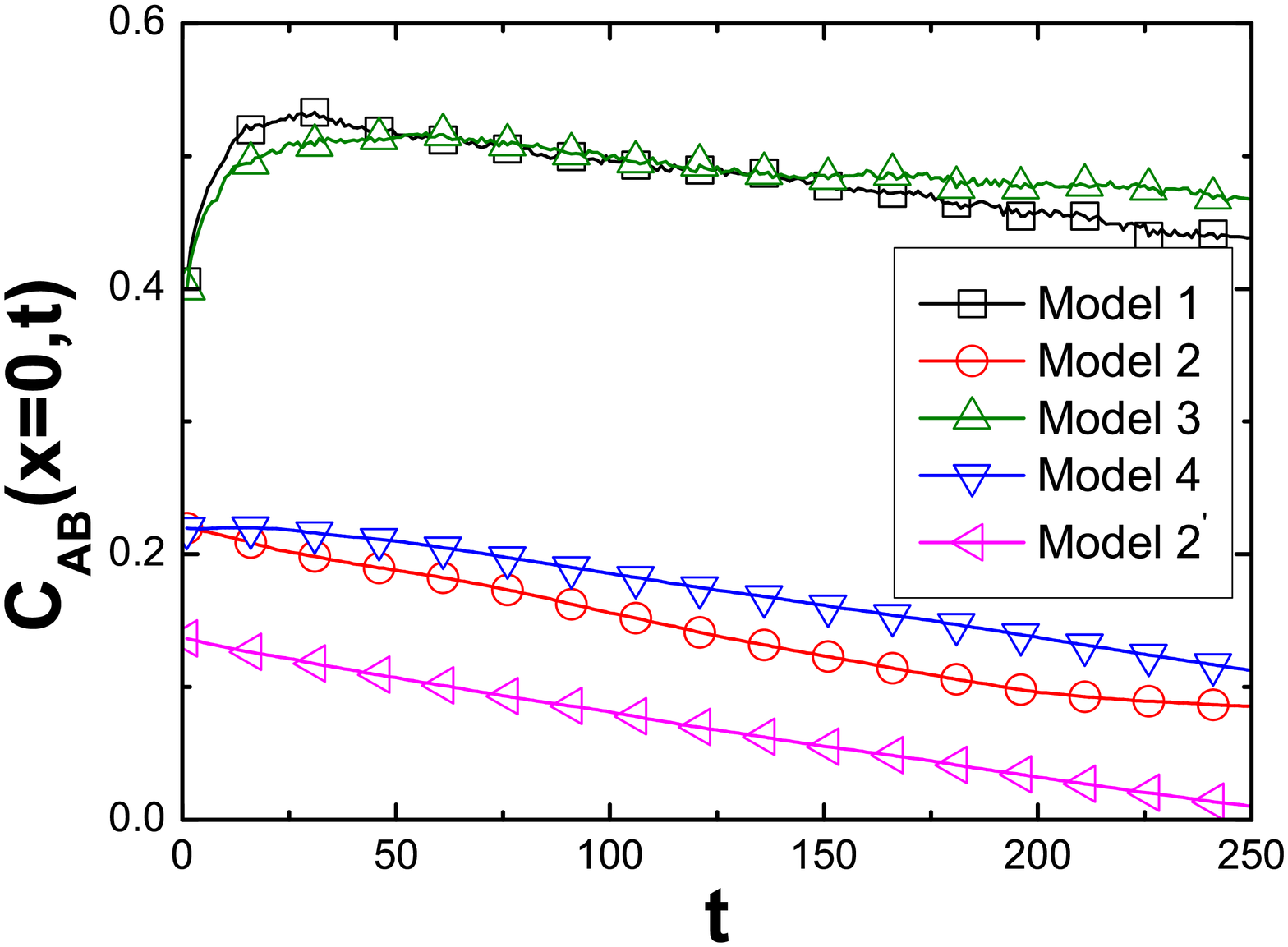}} \hskip -0.5cm
\subfloat[]{\label{fig15:cmpab}
\includegraphics[width=0.24\textwidth]{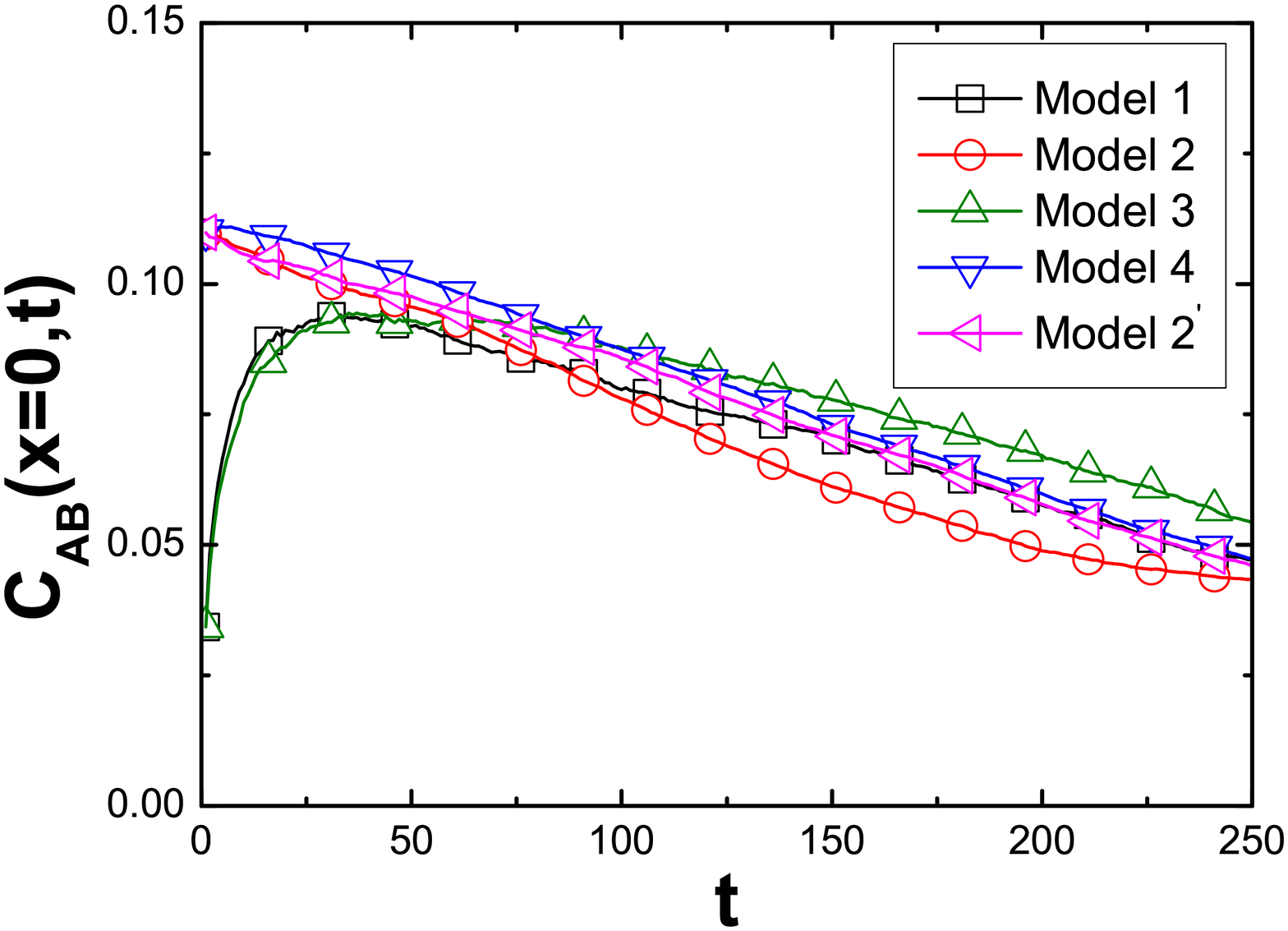}}
\caption{{\it (Color online.)} Time evolution for the on-site (a)
  autocorrelation $C_{AA}(0,t)$ and (b) cross-correlation function
  $C_{AB}(0,t)$ in one-dimensional RPS model variants with 500 sites, averaged
  over 1000 simulation runs.
  Shown are the results for model variants 1--4 with initial densities
  $a(0) = b(0) = c(0) = 1/3$; model 2' refers to a system with site occupancy
  restriction 1 and lower particle density $a(0) = b(0) = c(0) = 1/4$.}
\end{figure}

The population density signal Fourier transform $|a(f)|$, shown for species $A$
in Fig.~\ref{fig13:1dsift}, confirms the absence of any population oscillations
through the absence of any peak at nonzero frequency $f$, and the width of the
peak at $f = 0$ reflects the decay time to the stationary extinction state.
As in two dimensions, we observe very little effect of either site occupation
number restrictions or spatial variability of the reaction rates on the Fourier
signal, compare Figs.~\ref{fig2:2dftsa} and \ref{fig9:asdisf}.
We have also measured the mean single-species domain size
$\langle \lambda(t) \rangle$ and investigated its growth with time $t$, shown
in Fig.~\ref{fig14:1ddoms}.
As was predicted in Refs.~\cite{Frache1,Frache2}, for the implementation with
site occupancy restriction (model 2) to at most a single particle per site, we
observe $\langle \lambda(t) \rangle \sim t^{3/4}$; we find the same asymptotic
growth law when arbitrarily many particles are allowed on each lattice site.
An algebraic decay of the number of domains was also reported in
Ref.~\cite{Tainaka1}.
The domain stability is further illustrated by the very slow temporal decay of
the on-site auto- and cross-correlation functions (see Figs.~\ref{fig15:cmpaa}
and \ref{fig15:cmpab}).
Notice that quenched spatial disorder in the reaction rates does not affect
the time evolution of the autocorrelation functions, in contrast with site
occupancy restrictions; here the results depend on the presence or absence of
empty sites, see Fig.~\ref{fig15:cmpaa}.
However, the cross-correlation functions in Fig.~\ref{fig15:cmpab} look
essentially indistinguishable for all these model variations.

\begin{figure}[!b]
\subfloat[]{\label{fig16:1dcaa}
\includegraphics[width=0.24\textwidth]{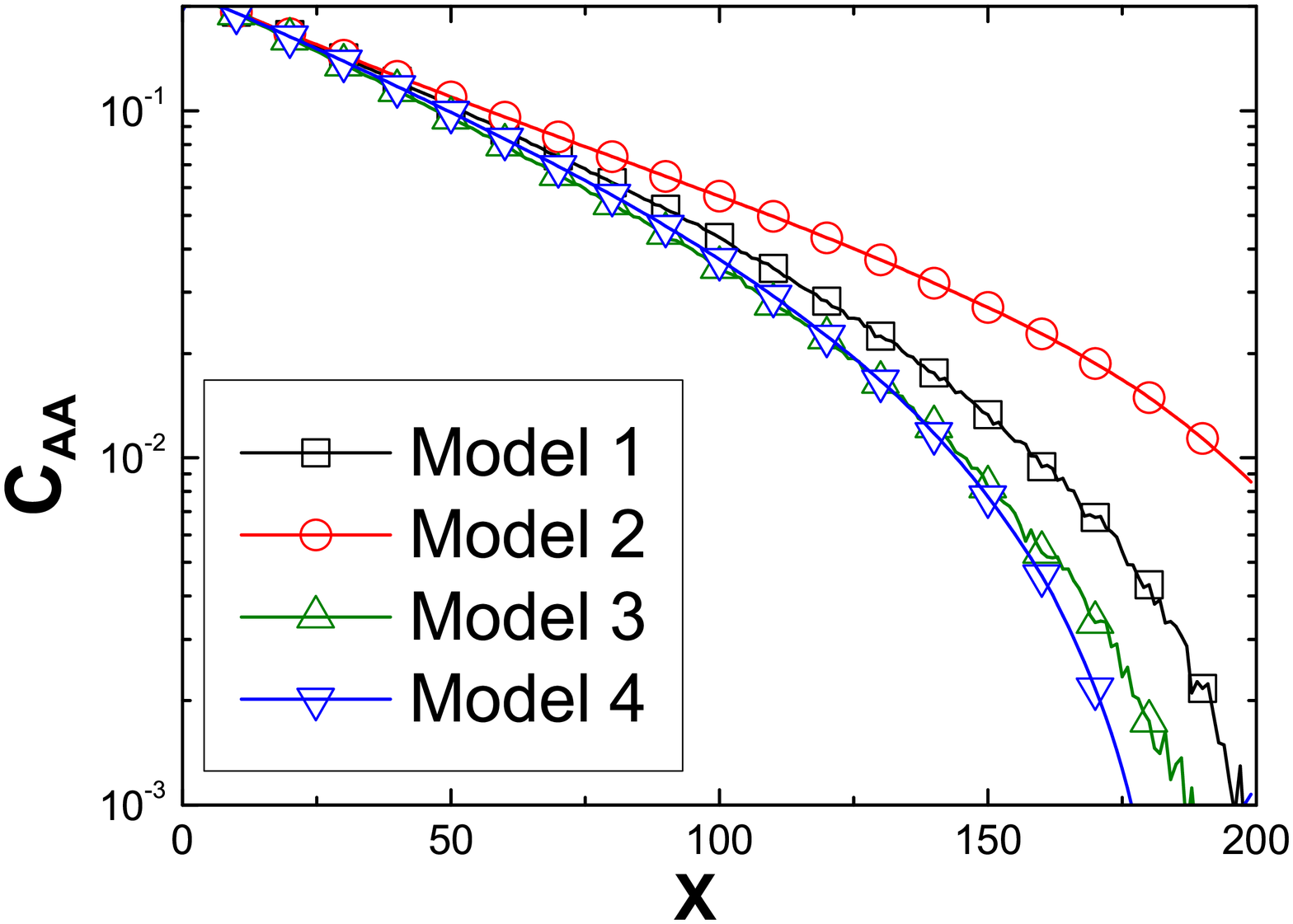}} \hskip -0.5cm
\subfloat[]{\label{fig16:1dcab}
\includegraphics[width=0.24\textwidth]{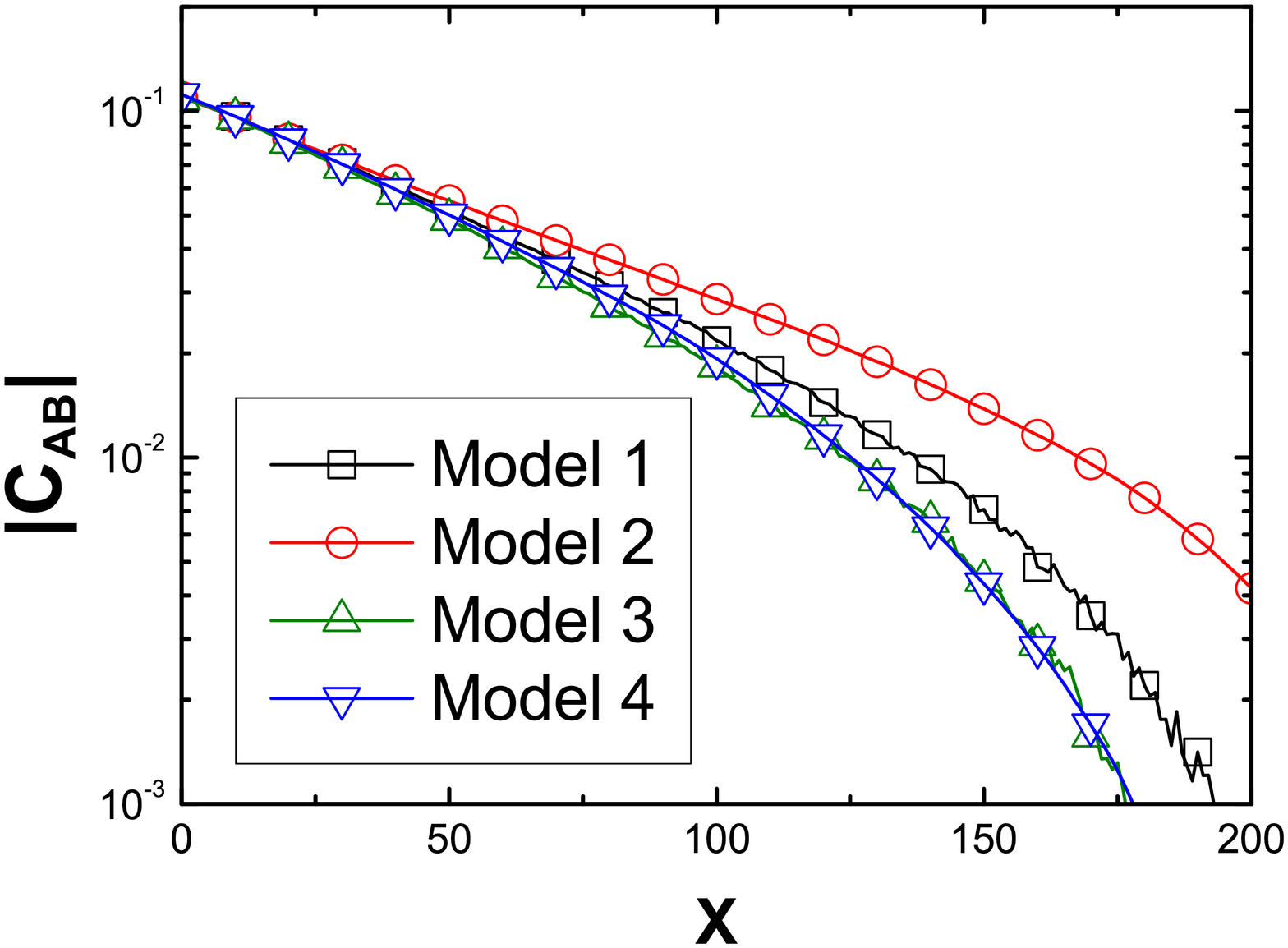}}
\caption{{\it (Color online.)} (a) Static autocorrelation functions $C_{AA}(x)$
  and (b) static cross-correlation functions $C_{AB}(x)$ (linear-$log_{10}$ 
  plots) measured at $t = 250$ MCS for the four RPS model variants described in 
  Table~\ref{models} on a one-dimensional lattice with 50000 sites, averaged 
  over 50 simulation runs.}
\end{figure}

Figures~\ref{fig16:1dcaa} and \ref{fig16:1dcab} respectively depict the
equal-time auto- and cross-correlation functions for the various model variants
listed in Table~\ref{models} obtained for a one-dimensional lattice with 50000
sites.
We observe exponential decay with similar large correlation lengths for all
model variants upto about 50 lattice sites, followed by a cutoff (which extends
to larger $x$ as time increases).

\begin{figure}[!t]
\subfloat[]{\label{fig17:1dex1}
\includegraphics[width=0.26\textwidth]{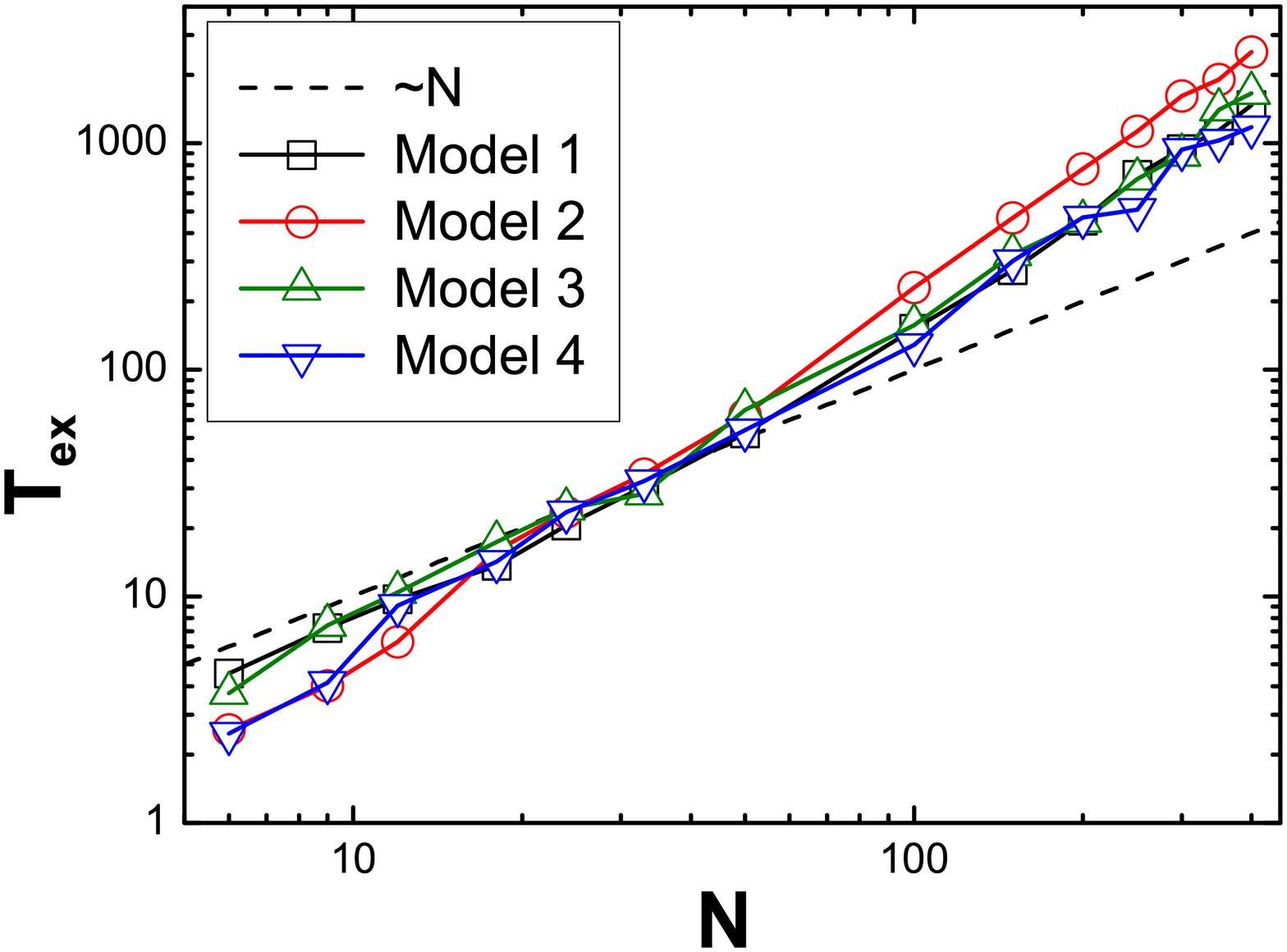}} \hskip -1cm
\subfloat[]{\label{fig17:exds2}
\includegraphics[width=0.26\textwidth]{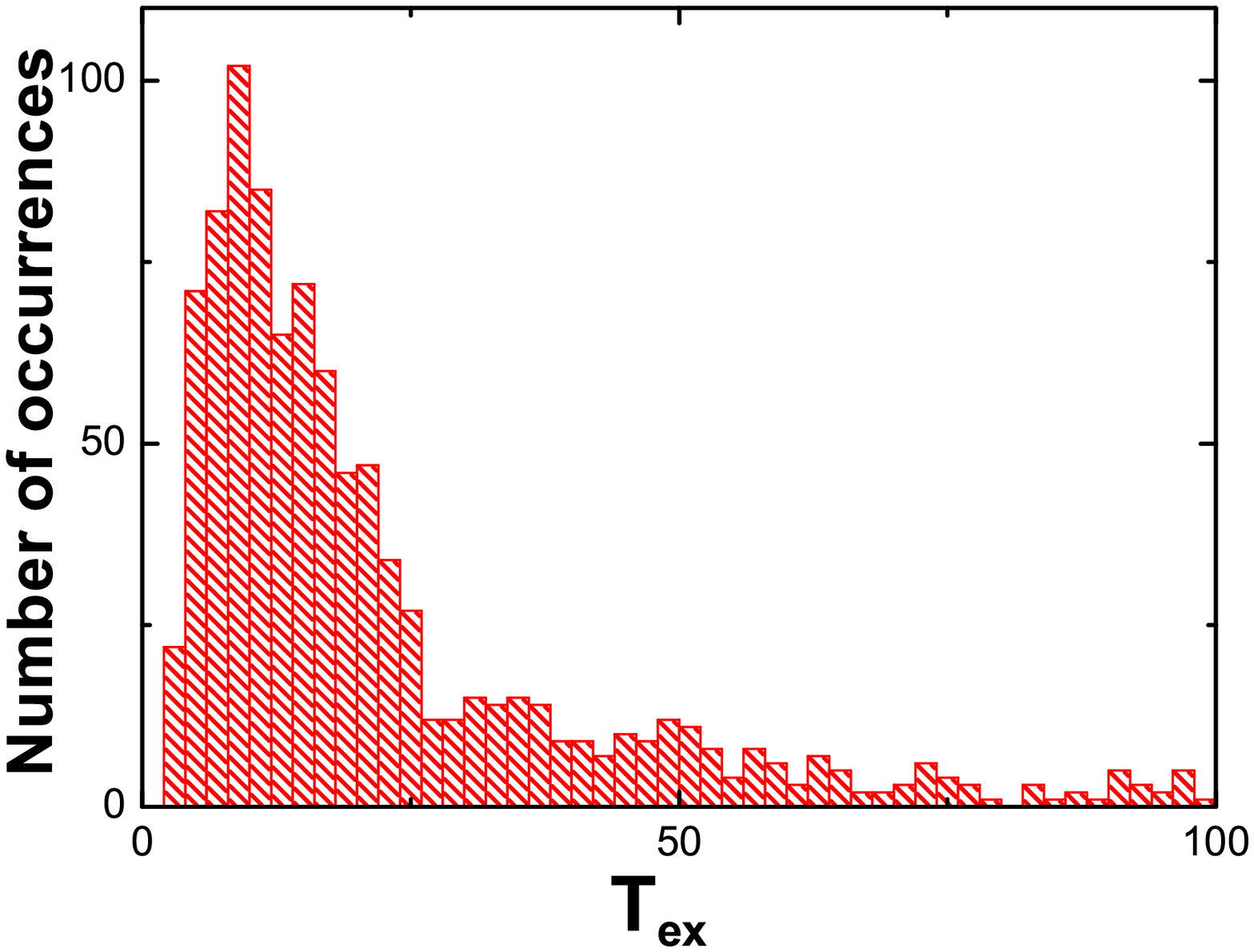}}
\caption{{\it (Color online.)} (a) Mean extinction time as function of lattice
  size $N$ ($log_{10}-log_{10}$ plot), obtained from averages over 1000 Monte 
  Carlo runs, in one dimension.
  (b) Histogram of the measured extinction times for model variant 2 with
  $N = 30$; compare with Figs.~\ref{fig5:exttwo} and \ref{fig5:exdis2}.}
\end{figure}

Finally, we investigate the mean extinction time as function of system size $N$
in one dimension.
As becomes apparent in Figs.~\ref{fig17:1dex1}, in all one-dimensional model
variants we have considered, within our (large) error bars the mean extinction
time appears to follow a power law $T_{\rm ex} \sim N^\gamma$, as proposed in
Refs.~\cite{Schutt,Reichen,Dobrienski,Parker}.
However, a best power-law fit yields variable effective exponents
$T_{\rm ex} \sim N^\gamma$ with $\gamma$ ranging from $\sim 1.5$ to $ \sim 1.8$
if we fit the data up to $N = 50$ or $N = 200$, respectively, rather than
$\gamma = 2$ \cite{Schutt} or $\gamma = 1$ \cite{Reichen,Dobrienski,Parker}.
Biasing the data towards smaller systems for which the statistical errors are
likely better controlled, our results may even be consistent with the
mean-field value $\gamma = 1$.
Note, however, that the extinction time distribution acquires even fatter tails
at large times than in two dimensions, see Fig.~\ref{fig17:exds2}, and rare
long survival events dominate the averages and induce large statistical
fluctuations.
The mean extinction time alone therefore poorly characterizes the extinction
kinetics.
When the reaction rates are chosen asymmetric, we have checked that only the
`weakest' species with the smallest predation rate survives, whereas the other
two species are driven to extinction~\cite{Frean,Berr}.

\section{Conclusion}
\label{conclu}

In this paper, we have studied the effects of finite carrying capacity and
spatial variability in the reaction rates on the dynamics of a class of spatial
rock--paper--scissors (RPS) models.
We have investigated the properties of several variants of the stochastic
four-state zero-sum RPS game (with conserved total particle number) on two- and
one-dimensional lattices with periodic boundary conditions.
In two dimensions, owing to the strict (local) conservation of the total
particle number, one does not observe the formation of spiral patterns; the
three species simply form small clusters.
In fact, spatial correlations are weak in the (quasi-stationary) coexistence
state, and the system is remarkably well described by the mean-field rate
equation approximation.
Typical extinction times scale exponentially with system size
\cite{Reichenbach4}, resulting in coexistence of all three species already on
moderately large lattices.
We find the characteristic initial oscillation frequency to be proportional to
the reaction rate and total particle density, as predicted by mean-field
theory.

We observe that neither site occupation number restrictions nor quenched
spatial disorder in the reaction rates markedly modify the populations'
temporal evolution, species density Fourier signals, or equal-time spatial
correlation functions.
This observation holds for models with symmetric as well as asymmetric reaction
rates, and even if spatial variability is introduced only for the competition
of one species pair.
On the basis of the mean-field results, this very weak disorder effect is a
consequence of the essentially {\em linear} dependence of the long-time
densities on the reaction rates $k$; averaging over a symmetric distribution
just yields the average.
In the two-species Lotka--Volterra model, instead both the asymptotic predator
and prey densities are inversely proportional to the predation rate, and
averaging over a distribution of the latter strongly biases towards small rate
values and large densities \cite{Ulrich}.
Both in the two- and cyclic three-species systems, spatially variable rates
induce stronger localization of the species clusters.

In one dimension, two species are driven towards extinction with the mean
extinction time $T_{\rm ex} \sim N^\gamma$, $\gamma \approx 1 \ldots 1.8$
(with large error bars), and only a single species survives.
The distribution of extinction times displays fat long-time tails.
We confirm that the single-species domains grow with the predicted power law
$\langle \lambda(t) \rangle \sim t^{3/4}$ in the models with
site restrictions \cite{Frache1,Frache2}, and we have obtained similar results
for the model variants with an infinite carrying capacity.
For asymmetric reaction rates, we have also checked that the `weakest' species
is the surviving one \cite{Frean,Berr}.

Our results demonstrate that the physical properties of cyclic RPS models are
quite robust, even quantitatively, with respect to modifications of their
`microscopic' model definitions and characterization.
This is in stark contrast with the related two-species Lotka--Volterra
predator-prey interaction model.
We believe the origin of this remarkable robustness lies in the comparatively
weaker prominence of stochastic fluctuations and spatial correlations in the
cyclic three-species system.
The robustness of the RPS models considered here implies that environmental
noise can be safely ignored and that their properties are essentially
independent of the carrying capacity.
It is worth emphasizing that this result is nontrivial; in terms of modeling
such systems, it notably implies that one has the freedom to consider strict
site restriction ($n_m = 1$) and hence simplify the numerical calculations, or
to set $n_m = \infty$ (no site restrictions) and thus facilitate the
mathematical treatment.
In fact, our study establishes that both `microscopic' model realizations are
essentially equivalent.
As it turns out, this conclusion also pertains to spatial May--Leonard models
\cite{MayLeonard,DurrettLevin}, as we shall report in detail elsewhere
\cite{HMT}.

\acknowledgments
This work is in part supported by Virginia Tech's Institute for Critical
Technology and Applied Science (ICTAS) through a Doctoral Scholarship.
We gratefully acknowledge inspiring discussions with Sven Dorosz, Michel
Pleimling, Matthew Raum, and Siddharth Venkat.


\begin{thebibliography}{99}

\bibitem{May}
  R. M. May, {\em Stability and Complexity in Model Ecosystems}
  (Cambridge University Press, Cambridge, 1974).
\bibitem{Maynard}
  J. Maynard Smith, {\em Models in Ecology}
  (Cambridge University Press, Cambridge, 1974).
\bibitem{Michod}
  R. E. Michod, {\em Darwinian Dynamics}
  (Princeton University Press, Princeton, 2000).
\bibitem{Sole}
  R. V. Sole and J. Basecompte, {\em Self-Organization in Complex Ecosystems}
  (Princeton University Press, Princeton, 2006).
\bibitem{Neal}
  D. Neal, {\em Introduction to Population Biology}
  (Cambridge University Press, Cambridge, 2004).
\bibitem{Lotka}
  A. J. Lotka, J. Amer. Chem. Soc. {\bf 42}, 1595 (1920).
\bibitem{Volterra}
  V. Volterra, Mem. Accad. Lincei. {\bf 2}, 31 (1926).
\bibitem{MaynardSmith}
  J. Maynard Smith, {\em Evolution and the Theory of Games}
  (Cambridge University Press, Cambridge, 1982).
\bibitem{Hofbauer}
  J. Hofbauer and K. Sigmund, {\em Evolutionary games and population dynamics}
  (Cambridge University Press, Cambridge, 1998).
\bibitem{Nowak}
  R. M. Nowak, {\em Evolutionary Dynamics}
  (Belknap Press, Cambridge(USA), 2006).
\bibitem{Szabo}
  G. Szab\'{o} and G. F\'{a}th, {\em Evolutionary games on graphs},
  Phys. Rep. {\bf 446}, 97 (2007).
\bibitem{Redner}
  D. Volovik, M. Mobilia, and S. Redner,
  Eur. Phys. Lett. {\bf 85}, 480003 (2009).
\bibitem{Mobilia}
  M. Mobilia, J. Theor. Biol. {\bf 264}, 1 (2010).
\bibitem{Sinervo}
  B. Sinervo and C.M. Lively, Nature {\bf 380}, 240 (1996).
\bibitem{Zamudio}
  K. R. Zamudio and B. Sinervo,
  Proc. Natl. Acad. Sci. U.S.A. {\bf 97}, 14427 (2000).
\bibitem{Kerr}
  B. Kerr, M. A. Riley, M. W. Feldman, and B. J. M. Bohannan,
  Nature {\bf 418}, 171 (2002).
\bibitem{Jackson}
  J. B. C. Jackson and L. W. Buss,
  Proc. Nat. Acad. Sci. U.S.A. {\bf 72}, 516 (1975);
  Am. Nat. {\bf 113}, 223 (1979).
\bibitem{Buss}
  L. W. Buss, Proc. Nat. Acad. Sci. U.S.A. {\bf 77}, 5355 (1980).
\bibitem{Burrows}
  M. T. Burrows and S. J. Hawkins, Mar. Ecol. Prog. Ser. {\bf 167}, 1 (1998).
\bibitem{Ifti}
  M. Ifti and B. Bergersen, Eur. Phys. J. E {\bf 10}, 241 (2003).
\bibitem{Reichen}
  T. Reichenbach, M. Mobilia, and E. Frey,
  Phys. Rev. E. {\bf 74}, 051907 (2006).
\bibitem{Frean}
  M. Frean and E. R. Abraham, Proc. R. Soc. Lond. B {\bf 268}, 1323 (2001).
\bibitem{Berr}
  M. Berr, T. Reichenbach, M. Schottenloher, and E. Frey,
  Phys. Rev. Lett. {\bf 102}, 048102 (2009).
\bibitem{Tainaka1}
  K. I. Tainaka, J. Phys. Soc. Jpn. {\bf 57}, 2588 (1988).
\bibitem{Frache1}
  L. Frachebourg, P. L. Krapivsky, and E. Ben-Naim,
  Phys. Rev. Lett. {\bf 77}, 2125 (1996).
\bibitem{Frache2}
  L. Frachebourg, P. L. Krapivsky, and E. Ben-Naim,
  Phys. Rev. E. {\bf 54}, 6186 (1996).
\bibitem{Provata}
  A. Provata, G. Nicolis, and F. Baras, J. Chem. Phys. {\bf 110}, 8361 (1999).
\bibitem{Pleimling}
  S. Venkat and M. Pleimling, Phys. Rev. E {\bf 81}, 021917 (2010).
\bibitem{Tainaka2}
  K. I. Tainaka, Phys. Lett. A {\bf 176}, 303 (1993).
\bibitem{Tainaka3}
  K. I. Tainaka, Phys. Rev. E {\bf 50}, 3401 (1994).
\bibitem{Szabo2}
  G. Szab\'{o} and A. Szolnoki, Phys. Rev. E {\bf 65}, 036115 (2002).
\bibitem{Perc}
  M. Perc, A. Szolnoki and G. Szabo, Phys. Rev. E {\bf 75}, 052102 (2007).
\bibitem{Reichenbach1}
  T. Reichenbach, M. Mobilia, and E. Frey, Nature {\bf 448}, 1046 (2007).
\bibitem{Reichenbach2}
  T. Reichenbach, M. Mobilia, and E. Frey,
  Phys. Rev. Lett. {\bf 99}, 238105 (2007).
\bibitem{Reichenbach3}
  T. Reichenbach, M. Mobilia, and E. Frey,
  {\em Banach Center Publications} Vol.{\bf 80}, 259 (2008).
\bibitem{Reichenbach4}
  T. Reichenbach, M. Mobilia, and E. Frey,
  J. Theor. Biol. {\bf 254}, 368 (2008).
\bibitem{Tsekouras}
  G. A. Tsekouras and A. Provata, Phys. Rev. E. {\bf 65}, 016204 (2001).
\bibitem{Matti}
  M. Peltom\"aki and M. Alava, Phys. Rev. E. {\bf 78}, 031906 (2008).
\bibitem{Dobrienski}
  A. Dobrienski and E. Frey, e-print {\tt arXiv:1001.5235}.
\bibitem{Ivan}
  M. Mobilia, I. T. Georgiev, and U. C. T\"auber,
  J. Stat. Phys. {\bf 128}, 447 (2007); and references therein.
\bibitem{Mark}
  M. J. Washenberger, M. Mobilia, and U. C. T\"auber,
  J. Phys.: Condens. Matter. {\bf 19}, 065139 (2007).
\bibitem{Ulrich}
  U. Dobramysl and U. C. T\"auber, Phys. Rev. Lett. {\bf 101}, 258102 (2008).
\bibitem{Hubbell}
  S. P. Hubbell,
  {\em The Unified Neutral Theory of Biodiversity and Biogeography}
  (Princeton University Press, Princeton, 2001).
\bibitem{Hanski}
  I. A. Hanski and O. E. Gagiotti (Eds.),
  {\em Ecology, genetics, and evolution of metapopulations}
  (Elsevier Academic Press, 2004).
\bibitem{HMT}
  Q. He, M. Mobilia, and U.~C. T\"auber, {\em in preparation}.
\bibitem{MayLeonard}
  R. M. May and W. Leonard, SIAM J. Appl. Math. {\bf 29}, 243 (1975).
\bibitem{DurrettLevin}
  R. Durrett and S. Levin, J. Theor. Biol. {\bf 185}, 165 (1997);
  Theor. Pop. Biol. {\bf 53}, 30 (1998).
\bibitem{Schutt}
  M. Sch\"utt and J.C. Claussen, e-print {\tt arXiv:1003.2427}.
\bibitem{Parker}
  M. Parker and A. Kamenev, Phys. Rev. E {\bf 80}, 021129 (2009).

\end{thebibliography}
\end{document}